\documentclass[aps,tightenlines,showpacs,nofootinbib]{revtex4}

\usepackage{graphicx}  %
\usepackage{amsmath}
\usepackage{bm}  %
\usepackage{empheq}

\newcommand*\widefbox[1]{\fbox{\hspace{2em}#1\hspace{2em}}}

\newcommand{\bea}{\begin{eqnarray}}
\newcommand{\eea}{\end{eqnarray}}
\newcommand{\beq}{\begin{equation}}
\newcommand{\eeq}{\end{equation}}
\newcommand{\bqa}{\begin{eqnarray}}
\newcommand{\eqa}{\end{eqnarray}}

\def\mqo2{{\!\!\!}}

\begin{document}

\title{
Galilean-Invariant XEFT}
\author{Eric Braaten}
\affiliation{Department of Physics,
         The Ohio State University, Columbus, OH\ 43210, USA\\}
\date{\today}

\begin{abstract}
XEFT is a low-energy effective field theory for charm mesons and pions
that provides a systematically improvable description of the 
$X(3872)$ resonance.
A Galilean-invariant formulation of XEFT 
is introduced to exploit the fact that mass is very nearly 
conserved in the transition $D^{*0} \to D^0 \pi^0$.
The transitions $D^{*0} \to D^0 \pi^0$
and $X  \to D^0 \bar D^0 \pi^0$ are described explicitly in XEFT.
The effects of the decay $D^{*0} \to D^0 \gamma$
and of short-distance decay modes of the $X(3872)$,
such as $J/\psi\, \pi^+ \pi^-$, can be taken into account
by using complex on-shell renormalization schemes for the
$D^{*0}$ propagator and for the $D^{*0} \bar D^0$ propagator 
in which the positions of their complex poles are specified.  
Galilean-invariant XEFT is used to calculate the 
$D^{*0} \bar D^0$ scattering length to next-to-leading order.
Galilean invariance ensures the cancellation of ultraviolet divergences
without the need for truncating an expansion in powers 
of the ratio of the pion and charm meson masses.
\end{abstract}

\smallskip
\maketitle

\section{Introduction}
\label{sec:intro}

The surprising discovery of the $X(3872)$ by the Belle Collaboration 
in 2003 \cite{Choi:2003ue} marked the beginning of a renaissance 
in quarkonium spectroscopy \cite{Olsen:2014qna}.
Dozens of new mesons whose constituents include a heavy quark 
and antiquark and with mass above the open-heavy flavor threshold
have been observed.  They are collectively referred to as 
$XYZ$ mesons.  Some of the $XYZ$ mesons are electrically charged
and therefore must be tetraquark mesons whose constituents 
also include a light quark and antiquark.
The pattern of the observed $XYZ$ mesons remains unexplained.
In particular, the relation between the $X(3872)$ 
and the other $XYZ$ mesons is still not understood.

The discovery decay mode $J/\psi\,\pi^+ \pi^-$ of the $X(3872)$
implies that its constituents must include a charm quark and antiquark.
However the $X(3872)$ has properties inconsistent with conventional 
charmonium, including comparable branching fractions 
into decay modes with isospin 0 and isospin 1.
The $J^{PC}$ quantum numbers of the $X(3872)$ were finally 
established by the LHCb Collaboration in 2013 
to be $1^{++}$ \cite{Aaij:2013zoa}.
Its mass is extremely close to the threshold for the pair 
of charm mesons $D^{*0} \bar D^0$.
By combining precise measurements of the mass $M_X$ of the $X(3872)$ 
in the $J/\psi\,\pi^+ \pi^-$ channel with precise measurements 
of the masses $M_*$ and $M$ of the $D^{*0}$ and $D^0$,
the difference $\delta_X$ between the $D^{*0} \bar D^0$ threshold  and the mass
has been determined to be 
\beq	
\delta_X  \equiv (M_* + M) - M_X =  0.11 \pm 0.23~{\rm MeV}.
\label{EX-exp}
\eeq

The quantum numbers $1^{++}$ of the $X(3872)$ imply that it has an
S-wave coupling to the charm meson pairs 
$D^{*0} \bar D^0$ and $D^0 \bar D^{*0}$.
Given the small value of $\delta_X$,
the universality of near-threshold S-wave resonances implies 
that the $X(3872)$ must be a bound state
(if $\delta_X > 0$) or a virtual state (if $\delta_X < 0$) 
whose constituents are the $C=+$ superposition 
$D^{*0} \bar D^0 + D^0 \bar D^{*0}$ \cite{Braaten:2003he}.
The observation of $X(3872)$ in hadron collisions strongly  suggests
that it is a bound state (like the deuteron)
rather than a virtual state (like the dineutron).
One universal property of S-wave near-threshold 
bound states is that the mean separation $\langle r \rangle_X$
of the constituents is determined by the binding energy $\delta_X$:
$\langle r \rangle_X= (8 \mu \delta_X)^{1/2}$,
where $\mu$ is the reduced mass of $D^{*0} \bar D^0$ \cite{Braaten:2004rn}.
Given the binding energy $\delta_X = 0.11^{-0.11}_{+0.23}$~MeV,
the mean separation of the charm mesons is predicted to be
$6.8^{+\infty}_{-3.9}$~fm.  Thus the size of the $X(3872)$ is 
comparable to that of the largest nuclei.

The universal properties of S-wave near-threshold 
bound states imply that most of the probability of the $X(3872)$
is in a molecular component consisting of well-separated
charm mesons $D^{*0} \bar D^0$.  (From now on, 
the equally probable $D^0 \bar D^{*0}$ component
will usually not be mentioned explicitly.)
At short distances, the wavefunction of the $X(3872)$ can have other
components with smaller probabilities. 
One possible component is the $1^{++}$ P-wave charmonium state 
$\chi_{c1}(2P)$, whose constituents are $c \bar c$.
Another possibility is an isospin-0 tetraquark with constituents
$c q \bar c \bar q$, where $q$ is a light $u$ or $d$ quark.
It could be a compact tetraquark, it could have substructure
consisting of the diquark clusters $c q$ and $\bar c \bar q$, 
or it could have substructure consisting of color-singlet clusters, 
such as the pair of mesons $J/\psi\, \omega$.
There are also hexaquark components of the  wavefunction 
with constituents $c q q \bar c \bar q \bar q$.  
One such component that is particularly important is $D^0 \bar D^0 \pi^0$, 
because the constituent $D^{*0}$ in the dominant component 
of the wavefunction can decay into $D^0 \pi^0$.   
In considering the various possible components of the $X(3872)$
wavefunction, it is essential to take into account their 
couplings to $D^{*0} \bar D^0$ and the resonant interactions 
between the charm mesons.

The interplay between the $D^{*0} \bar D^0$, $D^0 \bar D^0 \pi^0$,
and other components of the wavefunction can be treated
systematically using an effective field theory in which
the $D^{*0}$, $\bar D^0$,  and $\pi^0$ are explicit degrees of freedom.
Such an effective field theory has been developed by
Fleming, Kusunoki, Mehen, and van Kolck 
and named XEFT \cite{Fleming:2007rp}.
In its simplest form, XEFT is a nonrelativistic field theory
for $D^{*0}$ and $D^0$, their antiparticles, and $\pi^0$.
It is straightforward to extend XEFT
to include the charged charm mesons and the charged pions.
In XEFT, the interactions between $D^{*0}$ and $\bar D^0$ 
must be treated 
nonperturbatively in order to generate a bound state 
that can be identified with the $X(3872)$.
Fleming et al.\ showed that pion-exchange interactions
can be treated perturbatively along with the range corrections 
to $D^* \bar D$ interactions \cite{Fleming:2007rp}.
We denote the masses of $D^{*0}$, $D^0$, and $\pi^0$
by $M_*$, $M$, and $m$, respectively. 
We denote the difference between the $D^{*0}$ mass
and the sum of the $D^0$ and $\pi^0$ masses
by $\delta$:  
\beq	
\delta  \equiv M_* - (M+m) \approx 7.14 \pm 0.07~{\rm MeV}.
\label{delta-exp}
\eeq
The power counting of XEFT is defined by taking
external momenta, the binding momentum scale $\sqrt{M \delta_X}$ of the $X(3872)$,
and the momentum scale $\sqrt{m \delta}$ of the pion
to all be low momentum scales of order $Q$.
The low energy scales include the binding energy $\delta_X$ of the $X(3872)$,
the kinetic energy scale $\delta$ for a pion, and the kinetic energy scale 
$m \delta/M$ for a charm meson.
The high momentum or energy scales include $m$, $M$, and
$4 \pi f_\pi$, where $f_\pi$ is the pion decay constant.
Amplitudes are calculated as systematic expansions 
in powers of $Q$ divided by a high momentum or energy scale.
In the original paper on XEFT by Fleming et al.,
the momentum distributions in the decay of $X(3872)$ 
to $D^0 \bar D^0 \pi^0$ were calculated to next-to-leading order 
(NLO) in the XEFT power counting and to leading order in an
expansion in powers of $m/M$ \cite{Fleming:2007rp}.

There have been a number of subsequent applications of XEFT.
Fleming and Mehen  applied XEFT at leading order (LO) to decays of the 
$X = X(3872)$ into the P-wave charmonium state $\chi_{cJ}$ 
plus one or two pions \cite{Fleming:2008yn,Fleming:2011xa}.
Mehen and Springer  applied XEFT at LO to the radiative decays 
$X \to \psi(2S)\gamma$ and 
$\psi(4040) \to X \gamma$ \cite{Mehen:2011ds}.
Margaryan and Springer applied XEFT at LO to the decay
$\psi(4160) \to X \gamma$ \cite{Margaryan:2013tta}.
Recently, Jansen, Hammer and Jia used XEFT to 
determine the dependence of the binding energy of $X$
on the light quark masses \cite{Jansen:2013cba}. 
They also calculated the $D^{*0} \bar D^0$ scattering length
to NLO  and to leading order in $m/M$.
The applications of XEFT are not limited to the $D \bar D \pi$ sector,
which also includes $D^* \bar D$ and $D \bar D^*$ states.
Canham, Hammer, and Springer pointed out that XEFT 
can be applied to the  $D D \bar D \pi$ and 
$D D \bar D \pi \pi$ sectors \cite{Canham:2009zq}.  They calculated 
the S-wave phase shifts for low-energy scattering of $D$ or $D^*$ 
from $X$ at LO.
Braaten, Hammer, and Mehen  pointed out that XEFT can be applied 
to the $D \bar D \pi \pi$ sector, which also includes
$D^* \bar D^*$, $D^* \bar D \pi$,  and
$D \bar D^* \pi$ states \cite{Braaten:2010mg}.
They calculated the low-energy cross sections for 
elastic $\pi^+ X$ scattering and for the break-up reaction 
$\pi^+ X \to D^{*+} \bar D^0$ at LO.

We will refer to the formulation
of XEFT presented  in Ref.~\cite{Fleming:2007rp} as {\it original XEFT}.
There are various problems with original XEFT that present obstacles 
to accurate quantitative predictions.  
Some of these problems are due to its formulation as a nonrelativistic field theory 
of the charm mesons and $\pi^0$ that is not Galilean invariant.  
One consequence of 
the lack of Galilean invariance is that a preferred frame
(such as the center-of-momentum frame) must 
be specified either explicitly or implicitly in any calculation.  
Another consequence of the lack of Galilean invariance
is that ultraviolet divergences are much less constrained.
The explicit counterterms  in the Lagrangian
for original XEFT  in Ref.~\cite{Fleming:2007rp}
are sufficient to eliminate ultraviolet divergences in NLO
calculations only if results are expanded in powers 
of $m/M$ and then truncated at a sufficiently low order.
This truncation provides a limit on the accuracy.
Although the mass ratio $m/M = 0.072$ is very small,
the expansion is actually in powers of the square root of the mass ratio, 
which is 0.27.
The alternative of adding additional counterterms to the Lagrangian
to cancel the ultraviolet divergences would
introduce additional parameters that would have to be determined 
phenomenologically.
Another problem with original XEFT is that the renormalization scheme 
made it difficult to take into account some decays into final states 
with momenta  too large to be treated explicitly in XEFT.
These decays are $D^{*0} \to D^0 \gamma$, 
which accounts for about a third of the full width of the $D^{*0}$,
and all decay modes of the $X(3872)$ 
other than $D^0 \bar D^0 \pi^0$.

Alhakami and Birse have recently proposed an alternative power counting
for XEFT \cite{Alhakami:2015uea}.
In their power counting, $m$ and $M_* - M$
are treated as small energy scales of order $Q$, and 
$\delta$ is treated as a tiny energy scale of order $Q^2/M$.
The ratios $\delta/m = 0.051$ and $m/M = 0.072$
are both of order $Q/M$.
This power counting scheme was not implemented at the lagrangian level,
but it provides an organizing principle for simplifying Feynman diagrams.
The  power counting of Ref.~\cite{Alhakami:2015uea}
makes the expansion in powers of $m/M$ systematic.
The formulation in Ref.~\cite{Alhakami:2015uea} 
is convenient for matching onto heavy hadron chiral perturbation theory,
which has been used in several applications of XEFT 
\cite{Fleming:2008yn,Fleming:2011xa,Mehen:2011ds,Margaryan:2013tta}.
It does not address the problems of frame dependence 
or of additional ultraviolet divergences.
With this new power counting, the decay  $D^{*0}$ into $D^0 \gamma$
is within the domain of applicability of XEFT.
However there are important decay modes of the $X(3872) $,
such as $J/\psi\, \pi^+\pi^-$,  
that remain outside the domain  of applicability of XEFT.

In this paper, we present a new formulation of XEFT 
that removes many of the obstacles 
to accurate quantitative calculations.
The new formulation is an effective field theory with a Galilean
symmetry that is motivated by the fact that mass is very nearly conserved 
in the transition $D^{*0} \to D^0 \pi^0$.
The Galilean symmetry solves the problem of frame dependence 
and it dramatically simplifies ultraviolet divergences.
To take into account the decay $D^{*0} \to D^0 \gamma$
and decay modes of the $X(3872)$ other than $D^0 \bar D^0 \pi^0$,
a new renormalization scheme is introduced 
that is expressed in terms of  the complex  energies of $D^{*0}$
and $X(3872)$.
The new Galilean-invariant formulation of XEFT is illustrated by a 
calculation of the $D^{*0} \bar D^0$ scattering length
to NLO in the XEFT power counting.
The cancellation of ultraviolet divergences for arbitrary values of $m/M$
is verified explicitly.

\section{Galilean-invariant XEFT}
\label{sec:newXEFT}

In this section,
we introduce a new formulation of XEFT as a Galilean-invariant field theory.
We describe the changes in the Lagrangian for the original XEFT 
defined in Ref.~\cite{Fleming:2007rp} that are required for Galilean invariance.
We explain how the effects of the decay $D^{*0} \to D^0 \gamma$
can be taken into account through the complex rest energy of the $D^{*0}$.
We then write down the next-to-leading order Lagrangian for Galilean-invariant XEFT,
and give its Feynman rules.

\subsection{Galilean Invariance}
\label{sec:giXEFT}

A unique feature of the decay $D^{*0} \to D^0 \pi^0$
is that mass is very nearly conserved:
the sum of the masses of the $D^0$ and $\pi^0$ is only about 
3.5\% lower than the mass of the $D^{*0}$.
Galilean invariance is a possible space-time symmetry 
of a nonrelativistic theory that requires exact mass conservation \cite{Rosen:1972sh}.
The mass that must be conserved is the kinetic mass,
which is the mass that appears in the denominator of the 
kinetic energy.
The approximate conservation of mass in the 
transition $D^{*0} \to D^0 \pi^0$ strongly motivates
 a Galilean-invariant formulation of XEFT.
The Lagrangian for original XEFT, including all terms required to calculate 
the decay of $X(3872)$ into $D^0 \bar D^0 \pi^0$ to 
next-to-leading order (NLO), was written down in Ref.~\cite{Fleming:2007rp}.
We will describe how the terms in this Lagrangian must be modified to make 
them Galilean invariant. The NLO Lagrangian for 
Galilean-invariant XEFT will be written down in Section~\ref{sec:Lagrangian}.

We choose the kinetic masses of the $D^0$ and $\pi^0$ to be 
their physical masses $M$ and $m$, respectively.
Conservation of kinetic mass then requires the kinetic mass
of $D^{*0}$ to be $M+m$.  The difference $\delta$ between the mass 
of the $D^{*0}$ and its kinetic mass must be taken into account
through its rest energy.
Kinetic mass conservation
requires the following change in the Lagrangian 
for original XEFT defined in Ref.~\cite{Fleming:2007rp}:
\begin{itemize}
\item
In the kinetic term 
$\bm{D}^\dagger \cdot \nabla^2 \bm{D}/(2 m_{D^*})$
for the $D^{*0}$, its mass $m_{D^*}$  must be replaced by $M+m$.
\end{itemize}
A similar  change must  be made in the kinetic term 
for the $\bar D^{*0}$.
It will be convenient to introduce the reduced kinetic mass
$\mu$ for $D^{*0} \bar D^0$ and the reduced  mass
$\mu_\pi$ for $D^0 \pi^0$:
\begin{subequations}
\bea
\mu &\equiv& \frac{M(M+m)}{2M+m} = 965.0~{\rm MeV},
\label{mu*}
\\
\mu_\pi &\equiv& \frac{mM}{M+m = 125.87~{\rm MeV}}.
\label{mupi}
\eea
\label{mumu}%
\end{subequations}
The ratio of these reduced masses is
\beq
r \equiv \frac{\mu_\pi}{\mu} = 0.1304.
\label{r-mu}
\eeq

Galilean invariance requires interaction terms to
be invariant under Galilean boosts, in which the momenta
of $\pi^0$, $D^0$, and $D^{*0}$ are boosted by a common
velocity vector $\bm{v}$ multiplied by their kinetic masses
$m$, $M$, and $M+m$, respectively. 
Galilean invariance requires three changes in the interaction terms
in the Lagrangian for original XEFT that given
in Ref.~\cite{Fleming:2007rp}:
\begin{itemize}
\item
In the pion interaction term 
$\bm{D}^\dagger \cdot D \bm{\nabla} \pi$,
the operator $\bm{\nabla}$ between $D$ and $\pi$ should be replaced by
$(M \overrightarrow{\bm{\nabla}} - m \overleftarrow{\bm{\nabla}})/(M+m)$.
\item
In the $\nabla^2$ $D^{*0} \bar D^0$ interaction term 
$(\bar D \bm{D})^\dagger \cdot 
\bar D (\overleftrightarrow{\nabla})^2 \bm{D}$,
the operator $(\overleftrightarrow{\nabla})^2$
between $\bar D$ and $\bm{D}$ should be replaced by
$4 (M \overrightarrow{\bm{\nabla}} - (M+m) \overleftarrow{\bm{\nabla}})^2
/(2M+m)^2$.
\item
In the interaction term
$(\bar D \bm{D})^\dagger \cdot 
\bar D D \bm{\nabla} \pi$ that describes the transition
of $D^0 \bar D^0 \pi^0$ to $D^{*0} \bar D^0$,
the operator $\bm{\nabla}$ between $\bar D D$ and $\pi$
should be replaced by
$(2 M \overrightarrow{\bm{\nabla}} - m \overleftarrow{\bm{\nabla}})/(2M+m)$.
\end{itemize}
Similar changes  must   be made in the 
hermitian conjugates and charge conjugates of these interaction terms.
The three modified interaction terms described above reduce to those of original XEFT
in the limit $m/M \to 0$.  The modified interaction terms ensure
the invariance of amplitudes under Galilean boosts.

Galilean invariance also  strongly constraints  ultraviolet divergences.
The changes above specify all terms in the Lagrangian for 
Galilean-invariant XEFT that are required to calculate the 
decay of $X(3872)$ into $D^0 \bar D^0 \pi^0$ 
to NLO in the XEFT power counting.
The results should be independent 
of the ultraviolet cutoff to all orders in $m/M$.
Without Galilean invariance,
there are three independent $\nabla^2$ $D^{*0} \bar D^0$ interaction terms 
and two independent $D^0 \bar D^0 \pi^0 \to D^{*0} \bar D^0$
interaction terms.  The  Lagrangian for original XEFT
defined in Ref.~\cite{Fleming:2007rp}
includes only one interaction term of each kind.
Cutoff independent results for the decay rate  of $X(3872)$
 into $D^0 \bar D^0 \pi^0$ at NLO were obtained
by also truncating the expansion in powers of $m/M$ 
 at leading order.

Galilean-invariant XEFT can be extended to include 
charged charm mesons and charged pions.
The $\pi^+$ must have the same kinetic mass $m$ as $\pi^0$.  
The $D^+$ must have the same kinetic mass $M$ as $D^0$.  
The $D^{*+}$ must have the same kinetic mass $M+m$ as $D^{*0}$.  
The difference between the mass and kinetic mass of a particle 
must be taken into account through its rest energy.

In a nonrelativistic effective field theory for charm mesons, 
one can impose a phase symmetry that guarantees the separate conservation 
of the number $N_c$ of charm quarks 
and the number $N_{\bar c}$ of charm antiquarks.
These quark numbers can be expressed in terms of meson numbers:
\begin{subequations}
\bea
N_c &=& N_{D^{*0}} + N_{D^0},
\label{Nc}
\\
N_{\bar c} &=& N_{\bar D^{*0}} + N_{\bar D^0}.
\label{Ncbar}
\eqa
\label{Nccbar}%
\end{subequations}
If one also considers charged charm mesons,
the charm quark number $N_c$ is the sum of the numbers of
$D^{*0}$, $D^{*+}$, $D^0$, and $D^+$.
In Galilean-invariant XEFT, the exact conservation of kinetic mass
in the transitions $D^{*0} \leftrightarrow D^0 \pi^0$ 
and $\bar D^{*0} \leftrightarrow \bar D^0 \pi^0$
provides motivation for introducing an additional phase symmetry
that guarantees the conservation of the {\it pion number} defined by
\beq	
N_\pi = N_{\pi^0} + N_{D^{*0}} + N_{\bar D^{*0}}.
\label{Npi}
\eeq
The name {\it pion number}  is appropriate 
since the $D^*$ can be interpreted as a P-wave $D \pi$ resonance.
In Galilean-invariant XEFT with charged charm mesons and pions,
the pion number is the sum of the numbers of $\pi^0$, $\pi^+$, $\pi^-$,
$D^{*0}$, $\bar D^{*0}$, $D^{*+}$, and $D^{*-}$.

\subsection{Choice of Rest Energies}
\label{sec:restE}

In a nonrelativistic field theory, deviations from
mass conservation in a reaction are taken into account 
through the rest energies of the particles involved.
The rest energy of a particle can be chosen to be its physical mass.
In a Galilean invariant theory, since the kinetic mass is conserved, 
the rest energy of a particle can equally well chosen as 
the difference between its physical mass and its kinetic mass.
If there are linear combinations of the particle numbers that are conserved,
field redefinitions can be used to set the rest energies of some of the 
particles to 0.
In XEFT, conservation of the charm quark number $N_c$ allows 
the rest energies of $D^0$ and $D^{*0}$ to be set to 0 and $M_*-M$, respectively.
Conservation of both $N_c$ and the pion number $N_\pi$ allows
the rest energies of $\pi^0$, $D^0$, and $D^{*0}$ to be set to 
0, 0,  and $\delta = M_*-M-m$, respectively.

The rest energy of a particle
can also be used to take into account its partial width
into decay modes that cannot be described explicitly in the effective field theory.
Such a partial width is taken into account through
a negative imaginary term in the rest energy.
The width of $D^{*0}$ is more than 5 orders of magnitude larger 
than the widths of $D^0$ and $\pi^0$, 
so the widths of $D^0$ and $\pi^0$ can be completely ignored.
The width of $D^{*0}$ come from its decay into $D^0 \pi^0$, 
which can be described explicitly in XEFT,
and from its decay into $D^0 \gamma$, whose
momenta are too large to be described explicitly.
The effect of the $D^0 \gamma$ decay mode can be taken into account
by including a term $-i \Gamma_{*0,\gamma}/2$
in the $D^{*0}$ rest energy, where $\Gamma_{*0,\gamma}$
is the partial width of $D^{*0}$ into $D^0 \gamma$,
whose branching fraction is approximately 38\%. 
If we use the complex on-shell renormalization scheme 
for the $D^{*0}$ propagator in which the position of the complex pole is specified,
the $D^{*0}$ rest energy also includes the term $-i\Gamma_{*0,\pi}/2$,
where $\Gamma_{*0,\pi}$ is the partial width of 
$D^{*0}$ into $D^0 \pi^0$,
whose branching fraction is approximately 62\%.
In this scheme, the complete imaginary part of the 
$D^{*0}$ rest energy is $-i\Gamma_{*0}/2$,
where $\Gamma_{*0} = \Gamma_{*0,\pi} + \Gamma_{*0,\gamma}$
is the full width of the $D^{*0}$.

In original XEFT,
the rest energies of $\pi^0$, $D^0$, and $D^{*0}$
were chosen to be $-\delta$, 0, and 0, respectively.
Setting the rest energy of $D^{*0}$ to 0 is inconvenient 
if we wish to take into account
the partial width of $D^{*0}$ into $D^0 \gamma$.
It is more convenient to choose the rest energies of  
$\pi^0$ and $D^0$ to be 0.  The rest energy of $D^{*0}$
is then $\delta-i \Gamma_{*0}/2$.
This requires the following change in the Lagrangian 
for original XEFT:
\begin{itemize}
\item
The rest-energy term $\delta \pi^\dagger \pi$ for the $\pi^0$
should be replaced by the rest-energy term
$-  (\delta-i \Gamma_{*0}/2) \bm{D}^\dagger \cdot \bm{D}$
for the $D^{*0}$.
\end{itemize}
There is a similar rest-energy term for the $\bar D^{*0}$.

If Galilean-invariant XEFT is extended to include 
charged charm mesons and charged pions,
mass differences and nonnegligible decay modes outside 
the domain of validity of the effective field theory can be taken into account
through the rest energies of the particles.
The $\pi^+-\pi^0$ mass difference 
can be taken into account through the rest energy of $\pi^+$.
The $D^+-D^0$ mass difference 
can be taken into account through the rest energy of $D^+$.
The difference between the $D^{*+}-D^0\pi^0$ mass difference
can be taken into account through the rest energy of $D^{*+}$.
The tiny partial width for the decay of $D^{*+}$ into $D^+ \gamma$
can also be taken into account through the rest energy of $D^{*+}$.

\subsection{NLO Lagrangian}
\label{sec:Lagrangian}

We can now write down the complete NLO Lagrangian for Galilean-invariant XEFT
for the neutral charm mesons and the $\pi^0$.
It includes only those terms  required to calculate the elastic scattering amplitude 
for $D^{*0} \bar D^0$ to NLO in the XEFT power counting.
The field for the $\pi^0$ is denoted by $\pi$. 
The fields for the $D^0$ and $\bar D^0$ are denoted by $D$ and $\bar D$. 
The fields for the $D^{*0}$ and $\bar D^{*0}$ are denoted by 
$\bm{D}$ and $\bar{\bm{D}}$. 
The Lagrangian is the sum of kinetic terms, interaction terms, and counterterms.

There are kinetic terms for the  $\pi^0$, $D^0$, $\bar D^0$, and $D^{*0}$, and $\bar D^{*0}$.
The kinetic terms for the $\pi^0$, $D^0$, and $D^{*0}$ are
\begin{subequations}
\bea
{\cal L}_{\pi^0} &=& \pi^\dagger \left( i \partial_0 + \nabla^2/(2m) \right) \pi,
\label{Lpi}
\\
{\cal L}_{D^0} &=& D^\dagger \left( i \partial_0 + \nabla^2/(2M) \right) D,
\label{LD}
\\
{\cal L}_{D^{*0}} &=& \bm{D}^\dagger \cdot \left( i \partial_0 + \nabla^2/(2(M+m)) \right) \bm{D}
-  (\delta-i \Gamma_{*0}/2) \bm{D}^\dagger \cdot \bm{D}.
\label{LD*}
\eqa
\label{Lkinetic}%
\end{subequations}
The kinetic terms for the $\bar D^0$ and $\bar D^{*0}$ can be obtained  from those for
$D^0$ and $D^{*0}$ by replacing $D$ by $\bar D$  and $\bm{D}$ by $\bar{\bm{D}}$.

The interaction terms consist of pion interactions, contact interactions, 
and $\nabla^2$ interactions.
The pion interaction terms for the transitions
$D^{*0} \leftrightarrow D^0 \pi^0$ are 
\beq	
{\cal L}_{D^{*0} \leftrightarrow D^0 \pi^0} =
\frac{g}{2\sqrt{m} f_\pi (M+m)} 
\left[ \bm{D}^\dagger \cdot 
(D  [M \overrightarrow{\bm{\nabla}} - m \overleftarrow{\bm{\nabla}}] \pi)
+ (D  [M \overrightarrow{\bm{\nabla}} - m \overleftarrow{\bm{\nabla}}] \pi)^\dagger
\cdot \bm{D} \right].
\label{Lintpi}
\eeq
The pion interaction terms for the transitions
$\bar D^{*0} \leftrightarrow \bar D^0 \pi^0$ are obtained by replacing 
$\bm{D}$ and $D$  by $\bar{\bm{D}}$ and $\bar D$.
The contact and $\nabla^2$ interactions are the same in the 
$D^{*0} \bar D^0 \to D^{*0} \bar D^0$, $D^0 \bar D^{*0} \leftrightarrow D^{*0} \bar D^0$, 
and $D^0 \bar D^{*0} \to D^0 \bar D^{*0}$ channels.  
The interaction terms in the $D^{*0} \bar D^0 \to D^{*0} \bar D^0$ channel are
\bea
{\cal L}_{D^{*0} \bar D^0 \to D^{*0} \bar D^0} &=& 
- \frac{C_0}{2}
(\bar D \bm{D})^\dagger \cdot (\bar D  \bm{D})
\nonumber \\
&& + \frac{C_2}{4(2M+m)^2}
\left[ (\bar D \bm{D})^\dagger \cdot 
(\bar D [M \overrightarrow{\bm{\nabla}} - (M+m) \overleftarrow{\bm{\nabla}}]^2 \bm{D}) 
\right.
\nonumber \\
&& \left. \hspace{3cm}
+ (\bar D [M \overrightarrow{\bm{\nabla}} - (M+m) \overleftarrow{\bm{\nabla}}]^2 \bm{D} )^\dagger \cdot (\bar D \bm{D}) \right] .
\label{LD*D}
\eqa
The interaction terms in the other three channels are  obtained by 
replacing $\bm{D}$ and $\bar D$  by $\bar{\bm{D}}$ and $D$
in the appropriate places.  
The coupling constants $g$ in Eq.~\eqref{Lintpi} and 
$C_0$ and $C_2$ in Eq.~\eqref{LD*D} are essentially the same as those 
in the Lagrangian for original XEFT defined in Ref.~\cite{Fleming:2007rp}.

To cancel ultraviolet divergences in Green functions at NLO,
it is also necessary to include counterterms in the Lagrangian.
The counterterms are the same in the 
$D^{*0} \bar D^0 \to D^{*0} \bar D^0$, $D^0 \bar D^{*0} \leftrightarrow D^{*0} \bar D^0$, 
and $D^0 \bar D^{*0} \to D^0 \bar D^{*0}$ channels.  
The counterterms in the $D^{*0} \bar D^0 \to D^{*0} \bar D^0$ channel are
\bea
{\cal L}_{\rm counterterm} =
- \frac{\delta C_0}{2} (\bar D \bm{D})^\dagger \cdot (\bar D  \bm{D})
- \frac{\delta D_0}{2}
(\bar D \bm{D})^\dagger \cdot i \partial_t (\bar D  \bm{D}).
\label{Lct}
\eqa
The counterterm with coefficient $\delta D_0$ is not needed to calculate 
on-shell quantities, because it can be reduced to the other  counterterm
by using field redefinitions.  It was omitted in Ref.~\cite{Fleming:2007rp},
because the momentum distribution for  $D^0 \bar D^0 \pi^0$ 
in the decay of the $X(3872)$ is an on-shell quantity.

\subsection{Feynman Rules}
\label{sec:Frules}

The terms in the NLO Lagrangian for Galilean-invariant XEFT
with neutral charm mesons and the $\pi^0$
are given in Eqs.~\eqref{Lkinetic}, \eqref{Lintpi}, \eqref{LD*D}, and \eqref{Lct}.
We will give the Feynman rules for this Lagrangian.
The Feynman rules that are actually used for calculations
beyond leading order will be enclosed in boxes.

In XEFT, the charm quark and antiquark numbers $N_c$ and $N_{\bar c}$ are conserved.
In Galilean-invariant XEFT, the pion number $N_\pi$ is also conserved.
These conservation laws can all be made manifest in the Feynman rules 
by appropriate notation for the propagators.  We use a dashed line 
for the pion propagator, a solid line for the $D$ and $\bar D$ propagators,
and a double line consisting of a solid and a dashed line
for the $D^*$ and $\bar D^*$ propagators. 
In the propagators for the mesons $D^*$ and $D$
that contain a charm quark, the solid line has a forward arrow. 
In the propagators for the mesons $\bar D^*$ and $\bar D$
that contain a charm antiquark, the solid line has a backward arrow.
It is sometimes convenient to omit the arrows on internal lines
of diagrams and to use the convention that there is an implied sum 
over the possible directions of the omitted arrows.

\begin{figure}[tb]
\centerline{\includegraphics*[height=3cm,angle=0,clip=true]{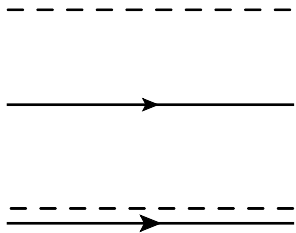}}
\vspace*{0.0cm}
\caption{
The  propagators for $\pi^0$, $D^0$, and $D^{*0}$
are represented by dashed, solid, and solid+dashed lines, respectively.  
The Feynman rules for these propagators are given in Eqs.~\eqref{prop}.
The propagators for $\bar D^0$ and $\bar D^{*0}$
look like those for $D^0$ and $D^{*0}$ with the arrows reversed.
}
\label{fig:prop}
\end{figure}

The propagators for the $\pi^0$, $D^0$, and $D^{*0}$ 
are illustrated in Fig.~\ref{fig:prop}.
The Feynman rules for the propagators of the $\pi^0$, 
the $D^0$ or $\bar D^0$,
and the $D^{*0}$ or $\bar D^{*0}$  are
\begin{subequations}
\begin{empheq}[box=\widefbox]{align}
& \frac{i}{p_0 - p^2/(2m) + i \epsilon},
\label{proppi}
\\
&\frac{i}{p_0 - p^2/(2M) + i \epsilon},
\label{propD}
\\
&\frac{i  \delta^{ij}}{p_0 - E_* -  p^2/(2(M+m)) },
\label{propD*}
\end{empheq}
\label{prop}%
\end{subequations}
where $p_0$ and $\bm{p}$ are the energy and momentum of the particle.
The propagator of the $D^{*0}$ is diagonal in its vector indices $i$ and $j$,
and its complex rest energy is
\beq	
E_* = \delta -i \Gamma_{*0}/2.
\label{vertex:E*}
\eeq
Since $E_*$ has a negative imaginary part,
an explicit $i \epsilon$ prescription is unnecessary
in the $D^{*0}$ propagator.

\begin{figure}[tb]
\centerline{\includegraphics*[height=3cm,angle=0,clip=true]{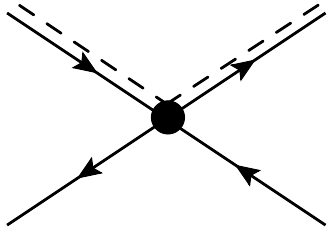}}
\vspace*{0.0cm}
\caption{
The contact interaction  vertex for $D^{*0} \bar D^0 \to D^{*0} \bar D^0$. 
The Feynman rule for this vertex is given in Eq.~\eqref{vertex:D*D}.
}
\label{fig:vertD*D}
\end{figure}

At LO in the XEFT power counting,
the only interaction that is required is a contact interaction
for $D^{*0} \bar D^0$  in the $C=+$ channel.  
The Feynman rule for the vertex is the same for the 
$D^{*0} \bar D^0 \to D^{*0} \bar D^0$, 
$D ^0 \bar D^{*0} \leftrightarrow D^{*0} \bar D^0$, 
and $D ^0 \bar D^{*0} \to D ^0 \bar D^{*0}$ contact interactions:
\beq	
\big( -i  C_0 /2 \big) \delta^{ij}.
\label{vertex:D*D}
\eeq
This vertex for $D^{*0} \bar D^0 \to D^{*0} \bar D^0$ is illustrated in Fig.~\ref{fig:vertD*D}.
The $C_0$ interaction must be treated nonperturbatively in XEFT.
The set of subdiagrams consisting of an arbitrary number of 
successive $C_0$ interactions can be 
summed to all orders analytically.  The Feynman rule for the resulting 
effective interaction is given later in Eq.~\eqref{amprule}.
In calculations beyond leading order in XEFT,
that effective vertex replaces the one in Eq.~\eqref{vertex:D*D}.
That is why the Feynman rule in Eq.~\eqref{vertex:D*D}
is not enclosed in a box.

\begin{figure}[tb]
\centerline{\includegraphics*[height=3cm,angle=0,clip=true]{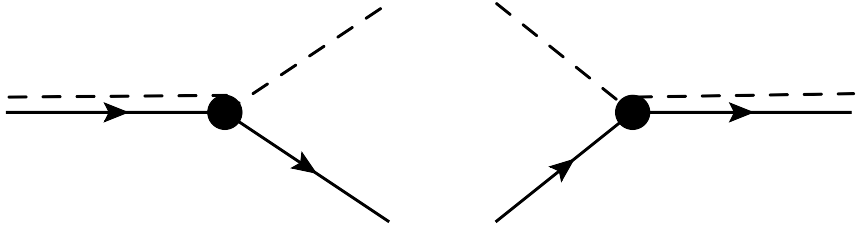}}
\vspace*{0.0cm}
\caption{
The vertices for the pionic transitions $D^{*0} \to D^0\pi^0$ and $D^0\pi^0 \to D^{*0}$. 
The Feynman rule for the $D^{*0} \to D^0\pi^0$ vertex is given in Eq.~\eqref{vertex:D*Dpi}.
}
\label{fig:vertpi}
\end{figure}

One of the interactions beyond LO in the XEFT power counting
is the $D^* \leftrightarrow D \pi$ transition.
The Feynman rules for the $D^{*0} \to D^0\pi^0$ and 
$\bar D^{*0} \to \bar D^0\pi^0$ vertices 
in Galilean-invariant XEFT are
\beq	
\boxed{
\frac{g}{2 \sqrt{m} f_\pi}
\frac{(M \bm{q} - m\bm{p})^i}{M+m},}
\label{vertex:D*Dpi}
\eeq
where $\bm{q}$ and $\bm{p}$ are the momenta of the 
outgoing $\pi^0$ and charm meson, respectively.
The vertices for $D^{*0} \to D^0\pi^0$ and 
$D^0\pi^0 \to D^{*0}$ are illustrated in Fig.~\ref{fig:vertpi}.
In the $D^0 \pi^0$  center-of-momentum frame defined by 
$\bm{p} + \bm{q} = 0$, the momentum-dependent factor
in Eq.~\eqref{vertex:D*Dpi} reduces to $q^i$,
which is the momentum-dependent factor in all frames in original XEFT.
The Feynman rules for the $D^0\pi^0 \to D^{*0} $ and 
$\bar D^0 \pi^0 \to \bar D^{*0}$ vertices differ 
from Eq.~\eqref{vertex:D*Dpi} by an overall minus sign
if $\bm{q}$ and $\bm{p}$ are the momenta of the 
incoming $\pi^0$ and charm meson, respectively.
A convenient way to implement the XEFT power counting is 
to assign orders in $g$ to the coupling constants 
of all other interaction terms.
A complete calculation then requires calculating
all diagrams to a given order in $g$.

\begin{figure}[tb]
\centerline{\includegraphics*[height=8cm,angle=0,clip=true]{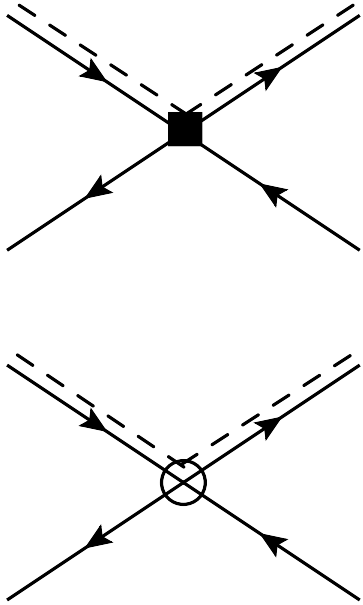}}
\vspace*{0.0cm}
\caption{
The $\nabla^2$ vertex and the counterterm vertex 
for $D^{*0} \bar D^0 \to D^{*0} \bar D^0$. 
The Feynman rules for these vertices are given in 
Eqs.~\eqref{vertex:grad^2} and \eqref{vertex:counterterm}, espectively.
}
\label{fig:vertnabla2}
\end{figure}

The other interaction at NLO in the XEFT power counting
is the $\nabla^2$  interaction
for $D^{*0} \bar D^0$ in the $C=+$ channel.
The Feynman rule for the vertex is the same for the 
$D^{*0} \bar D^0 \to D^{*0} \bar D^0$, 
$D ^0 \bar D^{*0} \leftrightarrow D^{*0} \bar D^0$, 
and $D ^0 \bar D^{*0} \to D ^0 \bar D^{*0}$ interactions.
In Galilean-invariant XEFT, the Feynman rule is
\beq	
\boxed{(-i  C_2/4)
\frac{( (M+m) \bm{p} - M \bm{p}_*)^2+ ( (M+m) \bm{p}' - M \bm{p}_*')^2}
       {(2M+m)^2} \delta^{ij},}
\label{vertex:grad^2}
\eeq
where $\bm{p}$ and $\bm{p}_*$ are the momenta of the 
incoming spin-0 and spin-1 charm mesons
and $\bm{p}'$ and $\bm{p}_*'$ are the momenta of the outgoing mesons.
The vertex for $D^{*0} \bar D^0 \to D^{*0} \bar D^0$ 
is illustrated in Fig.~\ref{fig:vertnabla2}.
In the $D^{*0} \bar D^0$  center-of-momentum frame defined by 
$\bm{p}_* + \bm{p} = \bm{p}_*' + \bm{p}' = 0$, the momentum-dependent factor
in Eq.~\eqref{vertex:grad^2} reduces to $p^2 + {p'}^2$.
In this frame, the Feynman rule coincides with that in original XEFT,
for which the momentum-dependent factor is 
$[(\bm{p} - \bm{p}_*)^2 + ( \bm{p}' -  \bm{p}_*')^2]/4$.
The coupling constant $C_2$ in Eq.~\eqref{vertex:grad^2} 
can be assigned order $g^2$.

In addition to these interaction vertices, it is also necessary to include
a counterterm vertex for $D^{*0} \bar D^0$ in the $C=+$ channel.
The vertex is the same for the 
$D^{*0} \bar D^0 \to D^{*0} \bar D^0$, 
$D ^0 \bar D^{*0} \leftrightarrow D^{*0} \bar D^0$, 
and $D ^0 \bar D^{*0} \to D ^0 \bar D^{*0}$ interactions:
\beq	
\boxed{\big(-i [\delta C_0 + \delta D_0 E]/2 \big) \delta^{ij},}
\label{vertex:counterterm}
\eeq
where $E$ is the total energy of the pair of charm mesons.
The counterterm vertex for $D^{*0} \bar D^0 \to D^{*0} \bar D^0$ 
is illustrated in  Fig.~\ref{fig:vertnabla2}.
The $\delta D_0$ term is not needed to calculate on-shell quantities
at NLO, but it is needed to cancel ultraviolet divergences in 
off-shell Green functions.
The counterterm coefficients  $\delta C_0$ and $\delta D_0$
in Eq.~\eqref{vertex:counterterm} can be assigned order $g^2$.

In the original paper on XEFT, an additional interaction term 
that produces a transition of $D^{*0} \bar D^0$ to $D^0 \bar D^0 \pi^0$
was written down explicitly \cite{Fleming:2007rp}.
The Feynman rule for its vertex
in Galilean-invariant XEFT is
\beq	
\boxed{\frac{B_1}{2\sqrt{m}}
\frac{( 2M \bm{q} - m (\bm{p} + \bm{\bar p}))^i}{2M+m}, }
\label{vertex:D*D-DDpi}
\eeq
where $\bm{q}$, $\bm{p}$, and $\bm{\bar p}$ are the outgoing momenta 
of the pion and the two spin-0 charm mesons.
In the $D^0 \bar D^0 \pi^0$  center-of-momentum frame defined by 
$\bm{p}  + \bm{\bar p}  + \bm{q} = 0$, the momentum-dependent factor
in Eq.~\eqref{vertex:D*D-DDpi} reduces to $q^i$,
which is the momentum-dependent factor in all frames in original XEFT.
The coupling constant $B_1$ in Eq.~\eqref{vertex:D*D-DDpi}
can be assigned order $g^3$.
This interaction term was needed in Ref.~\cite{Fleming:2007rp}
to calculate the decay of $X(3872)$ 
into $D^0 \bar D^0 \pi^0$ to NLO in the XEFT power counting
or, equivalently, to relative order $g^2$.  

In the $D^{*0}$ propagator in Eq.~\eqref{propD*},
a wavefunction renormalization factor $Z$ can be inserted in the numerator.
That factor can be made
completely arbitrary by inserting canceling factors of $Z^{-1/2}$
into the interaction vertices for each $D^{*0}$ or $\bar D^{*0}$ line.
The number of such factors in the interaction vertices in Eqs.~\eqref{vertex:D*D},
\eqref{vertex:D*Dpi}, \eqref{vertex:grad^2}, and \eqref{vertex:counterterm}
would be 2, 1, 2, and 2, respectively.
These factors can be absorbed into the coupling constants.

If we use dimensional regularization with $d$ spatial dimensions,
it is useful to introduce a renormalization scale $\Lambda$
to keep the dimensions of coupling constants the same as in
the physical dimension $d=3$. 
The interaction vertices in Eqs.~\eqref{vertex:D*D},
\eqref{vertex:D*Dpi}, \eqref{vertex:grad^2}, and \eqref{vertex:counterterm}
would be multiplied by $\Lambda$ raised to the powers 
$3-d$, $(3-d)/2$, $3-d$, and $3-d$, respectively.
In a Green function, the net effect of these powers of  $\Lambda$ is 
a factor of $\Lambda^{3-d}$ for every loop integral
and an overall multiplicative factor of $\Lambda^{3-d}$ raised to some power.
The factor of $\Lambda^{3-d}$ 
associated with a loop integral can be absorbed into its integration measure.
If the Green function is made finite by renormalization,
the overall multiplicative factor of $\Lambda^{3-d}$ raised to some power
can be simply be ignored, because it is equal to 1 
in the physical dimension $d=3$.

\section{Renormalization Prescriptions}
\label{sec:renorm}

In this section, we introduce renormalization prescriptions that can be used to 
take into account the effects of  transitions to states in which the momenta 
are too large to be described explicitly in XEFT.
We also calculate the transition amplitude for $D^{*0} \bar D^0 \to D^{*0} \bar D^0$
at leading order in XEFT.

\subsection{$\bm{D^*-D \pi}$ Coupling Constant}
\label{sec:D*Dpi}

One of the coupling constants in XEFT beyond leading order is 
the $D^*-D\pi$ coupling constant $g/\sqrt2 f_\pi$.
It could in principle be determined 
from the partial width of $D^{*0}$ into $D^0 \pi^0$:
\beq	
\Gamma_{*0,\pi} \equiv \Gamma[D^{*0} \to D^0 \pi^0] =
\left( \frac{g^2}{4m f_\pi^2} \right)
\frac{\mu_\pi}{3 \pi}  ( 2 \mu_\pi \delta)^{3/2},
\label{GammaD*Dpi0}
\eeq
where $\delta$ is the mass difference in Eq.~\eqref{delta-exp}
and $\mu_\pi$ is the $D \pi$ reduced mass 
defined in Eq.~\eqref{mupi}.
The branching fraction for $D^{*0} \to D^0 \pi^0$ 
has been measured to be $(62 \pm 3)\%$,
but the full width of $D^{*0}$ has not been measured.

By exploiting isospin symmetry,
the coupling constant $g/\sqrt2 f_\pi$ can also be determined 
from decays of the $D^{*+}$.
The full width of $D^{*+}$ has been recently measured
with high precision by the Babar Collaboration \cite{Lees:2013zna}:
$\Gamma[D^{*+}] = 83.3 \pm 1.2 \pm 1.4$~keV. 
The sum of the branching factions of $D^{*+}$ into $D^0 \pi^+$
and $D^+ \pi^0$  is $0.984 \pm 0.005$.
In Galilean-invariant XEFT,
the sum of the partial widths into $D^0 \pi^+$ and $D^+ \pi^0$ is
\beq	
\Gamma[D^{*+} \to D^0 \pi^+] +\Gamma[D^{*+} \to D^+ \pi^0] =
\frac{1}{3 \pi} \left( \frac{g^2}{4m f_\pi^2} \right) \mu_\pi
\left[ 2 ( 2 \mu_\pi \delta_{+0})^{3/2} + ( 2 \mu_\pi \delta_{++})^{3/2} \right],
\label{GammaD*Dpi+}
\eeq
where $\delta_{+0} =5.856\pm 0.002$~MeV is the 
$D^{*+}-D^0\pi^+$ mass difference, and $\delta_{++} =5.68\pm 0.08$~MeV 
is the $D^{*+}-D^+\pi^0$ mass difference.
The resulting value of the coupling constant $g$ is given by
\beq	
\frac{g^2}{4m f_\pi^2} =  (3.67 \pm 0.08) \times 10^{-8}
~{\rm MeV}^{-3}.
\label{gg:exp}
\eeq

The pion decay constant $f_\pi$ determines the scattering amplitude
for $\pi \pi$ scattering.
With the convention for $f_\pi$ used in Ref.~\cite{Fleming:2007rp}, 
its value is $f_\pi \approx 132$~MeV.
Given that value, the pion transition coupling constant
 in Eq.~\eqref{gg:exp} is determined to be $g \approx 0.59$.
A more appropriate dimensionless measure of the strength of the 
pion exchange interaction in XEFT is
\beq	
\frac{g^2\mu^2 \sqrt{ 2 \mu_\pi \delta}}{12 \pi m f_\pi^2} \approx 0.15,
\label{ggmu}
\eeq
where $\mu$ is the reduced kinetic mass of $D^{*0} \bar D^0$
defined in Eq.~\eqref{mu*}.

The determination of the values of $g$ and $f_\pi$ separately
is unnecessary in XEFT applied to the $D \bar D \pi$ sector
(which includes $D^* \bar D$),
because the number of pions in the system can only be 0 or 1.
In the $D \bar D \pi \pi$ sector (which includes $D^* \bar D^*$), 
the number of pions in the system can be 0, 1, or 2.
Since there can be contributions from
$\pi \pi$ scattering, the value of $f_\pi$ is needed 
at a sufficiently high order in the XEFT power counting.

\subsection{$\bm{D^{*0}}$ Propagator}
\label{sec:D*prop}

\begin{figure}[tb]
\centerline{\includegraphics*[height=3cm,angle=0,clip=true]{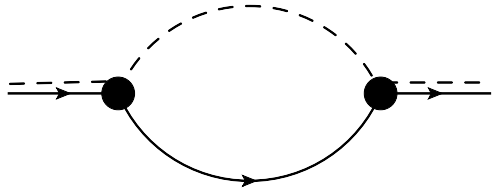}}
\vspace*{0.0cm}
\caption{
The one-loop (and only) $D^{*0}$ self-energy diagram in XEFT.
The diagram with external legs amputated can be expressed as
$-i  \Sigma(E_{\rm cm}) \delta^{ij}$.
}
\label{fig:D*selfenergy}
\end{figure}

The conservation of pion number in Galilean-invariant XEFT 
implies that the exact $D^{*0}$ propagator can be calculated analytically.
It can be obtained by summing a geometric series of the one-loop $D^{*0}$
self-energy diagram in Figure~\ref{fig:D*selfenergy}.
Galilean invariance implies that the propagator for a $D^{*0}$ 
with energy $p_0$ and momentum $\bm{p}$
is a function of the Galilean-invariant combination
\beq	
E_{\rm cm} = p_0 - \frac{p^2}{2(M+m)}.
\label{Ecm-D*}
\eeq
The exact $D^{*0}$ propagator can be expressed as 
\beq	
D_*^{ij}(p_0,p) = 
\frac{i [1 + \delta Z] \delta^{ij}}{E_{\rm cm} - (E_* +  \delta E_*) - \Sigma(E_{\rm cm})} ,
\label{propD*-exact}
\eeq
where $\Sigma(E_{\rm cm})$ is the $D^{*0}$ self-energy 
and $\delta E_*$ is a rest-energy counterterm. 
We have introduced a wavefunction renormalization factor 
$1 + \delta Z$ in the numerator.  It can be made completely arbitrary by absorbing
canceling factors of $[1 + \delta Z]^{-1/2}$ into the interaction vertices.
With dimensional regularization,
the $D^{*0}$ self-energy  is
\beq
\Sigma(E_{\rm cm}) =
- \mu_\pi \left( \frac{g^2}{4m f_\pi^2} \right) 
\frac{\Lambda^{3-d} \Gamma(-d/2)}{(4 \pi)^{d/2}}
\left[ e^{-i \pi} 2 \mu_\pi E_{\rm cm} \right]^{d/2} ,
\label{SigmaD*}
\eeq
where $\Lambda$ is the renormalization scale
and $\mu_\pi$ is the $D^0 \pi^0$ reduced mass in Eq.~\eqref{mupi}.
The self-energy has cubic and linear ultraviolet divergences that 
are manifested as single poles in $d$ and $d-2$, respectively.
The pole in $d-2$ is
\beq
\frac{1}{d-2} \lim_{d \to 2} (d-2) \Sigma(E_{\rm cm}) = 
\left( \frac{g^2}{4m f_\pi^2} \right)\frac{2 \mu_\pi^2 \Lambda}{3 \pi (d-2)} 
E_{\rm cm}.
\label{Sigma-PDS}
\eeq
In the physical dimension $d=3$, the self-energy is pure imaginary
for real positive $E_{\rm cm}$: 
\beq
\lim_{d \to 3} \Sigma(E_{\rm cm}) =
- i \left( \frac{g^2}{4m f_\pi^2} \right) 
\frac{\mu_\pi }{6 \pi} \left[ 2 \mu_\pi E_{\rm cm} \right]^{3/2} .
\label{Sigma:d=3}
\eeq
Its value at $E=\delta$ is $-i \Gamma_{*0,\pi}$/2,
where $\Gamma_{*0,\pi}$ 
is the partial width of $D^{*0}$ into $D^0 \pi^0$
given in Eq.~\eqref{GammaD*Dpi0}.

In the {\it minimal power-divergence subtraction} (PDS) 
renormalization scheme \cite{Kaplan:1998tg},
the $D^{*0}$ rest energy and the propagator counterterms  are
\begin{subequations}
\bea
E_{*,{\rm PDS}} &=& \delta,
\label{E*:PDS}
\\
\delta E_{*,{\rm PDS}} &=& 0,
\label{deltaE*:PDS}
\\
\delta Z_{\rm PDS} &=& 
-  \left( \frac{g^2}{4m f_\pi^2} \right) \frac{2 \mu_\pi^2 \Lambda}{3 \pi (d-2)} .
\label{Z*:PDS}
\eqa
\label{EdeltaEZ*:PDS}%
\end{subequations}
In the physical dimension $d=3$, 
the self-energy $\Sigma(E_{\rm cm})$ reduces to Eq.~\eqref{Sigma:d=3}.
The approximate position of the pole in $p_0$ of the 
$D^{*0}$ propagator in Eq.~\eqref{propD*-exact} is where $E_{\rm cm}$ 
is equal to the complex energy 
$\delta-i  \Gamma_{*0,\pi}/2$.
The PDS scheme does not take into account the decay $D^{*0} \to D^0 \gamma$. 

The pole in the energy $p_0$ for the physical $D^{*0}$ propagator
occurs when $E_{\rm cm}$ is equal to the complex energy 
$E_*= \delta-i\Gamma_{*0}/2$, where $\Gamma_{*0}$ 
is the full width of the $D^{*0}$.
We introduce the {\it complex on-shell} (COS) renormalization scheme 
for the $D^{*0}$ propagator in which its pole in $p_0$ is at the complex physical 
value and the residue of that pole is the same as at LO.
The $D^{*0}$ rest energy and the propagator counterterms are 
\begin{subequations}
\bea
E_{*} &=& \delta-i\Gamma_{*0}/2,
\label{E*}
\\
\delta E_{*} &=&- \Sigma(E_*),
\label{deltaE*}
\\
\delta Z &=&-  \Sigma'(E_*),
\label{Z*}
\eqa
\label{EdeltaEZ*}%
\end{subequations}
where $\Sigma(E)$ is the self-energy in Eq.~\eqref{SigmaD*}.
In the physical dimension $d=3$, 
$\delta E_*$ and $\delta Z$ in the COS scheme reduce to
\begin{subequations}
\bea
\lim_{d \to 3} \delta E_*  &=&
 -i \left( \frac{g^2}{4m f_\pi^2} \right) 
\frac{\mu_\pi}{6 \pi} \left[ 2 \mu_\pi E_* \right]^{3/2} ,
\label{deltaE*:d=3}
\\
\lim_{d \to 3} \delta Z &=& 
 i \left( \frac{g^2}{4m f_\pi^2} \right) 
\frac{\mu_\pi^2}{2 \pi} \left[ 2 \mu_\pi E_* \right]^{1/2} .
\label{Z*:d=3}
\eqa
\label{deltaEZ*:d=3}%
\end{subequations}
If we ignore the tiny difference between $E_* = \delta - i \Gamma_{*0}/2$ 
and $\delta$, the rest energy counterterm 
in Eq.~\eqref{deltaE*:d=3} reduces to 
$\delta E_* = - i \Gamma_{*0,\pi}/2$, where $ \Gamma_{*0,\pi}$
is the partial width of the $D^{*0}$ into $D^0 \pi^0$ in 
Eq.~\eqref{GammaD*Dpi0}.

\begin{figure}[tb]
\centerline{\includegraphics*[width=5cm,angle=0,clip=true]{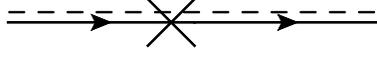}}
\vspace*{0.25cm}
\caption{
The $D^{*0}$ self-energy counterterm vertex.
The Feynman rule for this vertex at order $g^2$ in the complex on-shell scheme 
is given in Eq.~\eqref{D*counterterm}.
}
\label{fig:D*counterterm}
\end{figure}

Although Eq.~\eqref{propD*-exact} is the exact $D^{*0}$ propagator 
in Galilean-invariant XEFT, it is not practical to use this propagator
in loop diagrams, because the term in the denominator proportional 
to $E_{\rm cm}^{d/2}$  makes loop integrals 
more complicated to evaluate.  It is better to expand the 
exact propagator in Eq.~\eqref{propD*-exact} in powers of $g^2$:
\beq	
D_*^{ij}(p_0,p) = 
\frac{i   \delta^{ij}}{E_{\rm cm} - E_*}
\sum_{N=0}^\infty \left[ 
\frac{\Sigma(E_{\rm cm}) + \delta E_* + \delta Z (E_{\rm cm} - E_*)}
{[1 + \delta Z] (E_{\rm cm} - E_*) }  \right]^n.
\label{propD*-expand}
\eeq
The expansion generates a geometric series of propagator corrections.
The terms of order $g^2$ correspond to inserting into the propagator
the one-loop $D^{*0}$ self-energy subdiagram in Fig.~\ref{fig:D*counterterm}
and a $D^{*0}$ self-energy counterterm.
The Feynman rule  for the order-$g^2$ self-energy counterterm vertex
in the COS scheme is
\beq	
\boxed{+i \Big( \Sigma(E_*) 
+ \Sigma'(E_*) \big[ p_0 - p^2/(2(M+m)) - E_* \big] \Big)
\delta^{ij},}
\label{D*counterterm}
\eeq
where $p_0$ and $p$ are the energy and momentum  of the $D^{*0}$,
$i$ and $j$ are its vector indices,
and $\Sigma(E)$ is the self-energy in Eq.~\eqref{SigmaD*}.

The previous NLO calculations in XEFT were carried out in the 
original version of XEFT, which is not Galilean invariant.
The  $D^{*0}$ self-energy $\Sigma(p_0,p)$ is therefore a function 
of the two independent variables $p_0$ and $p$.
The kinetic mass of the $D^{*0}$ was set equal to its physical mass $M_*$.
The $D^{*0}$ rest energy was taken to be the real energy $\delta$,
although it was actually taken into account through the rest energy term for the $\pi^0$.
In the NLO calculation of the decay of $X(3872)$ into $D^0 \bar D^0 \pi^0$ 
in Ref.~\cite{Fleming:2007rp}, the renormalization of the $D^{*0}$ propagator 
was carried out using the minimal power divergence subtraction 
scheme in Eq.~\eqref{EdeltaEZ*:PDS}.
In the physical dimension $d = 3$, the counterterm $\delta Z_{\rm PDS}$ 
in Eq.~\eqref{Z*:PDS} reduces to the 
renormalization scale $\Lambda$ multiplied by a constant.
This term was cancelled by a kinetic counterterm for the $D^{*0}$.

In the NLO calculation of the $D^{*0} \bar D^0$ scattering length
In Ref.~\cite{Jansen:2013cba}, the complex energy of the $X(3872)$
and its partial width into $D^0 \bar D^0 \pi^0$
were calculated to NLO using original XEFT.
The calculation involved a two-loop diagram 
that has a one-loop $D^{*0}$ self-energy subdiagram 
and that diverges as $1/p$ as the relative momentum $p$
of $D^{*0} \bar D^0$ approaches 0.
The divergence was avoided by resumming to all orders terms 
proportional to $\Gamma_{*0,\pi}^{1/2}$
from one-loop $D^{*0}$ self-energy subdiagrams.
The resummation could have been implemented in such a way 
that the terms that were resummed had only integer powers of $\Gamma_{*0,\pi}$.
This is equivalent to the complex on-shell renormalization scheme 
for the $D^{*0}$ propagator in Eq.~\eqref{EdeltaEZ*},
but with $E_*$ replaced by $\delta - i \Gamma_{*0,\pi}/2$.
The imaginary part of $E_*$ is not equal to the physical value 
$- \Gamma_{*0}/2$, because the decay  $D^{*0} \to D^0 \gamma$ 
was not taken into account in Ref.~\cite{Jansen:2013cba}.

\subsection{$\bm{D^{*0} \bar{D^0}}$ Transition Amplitude}
\label{sec:transamp}

In XEFT, the conservation of pion number ensures that loop diagrams
in the $D \bar D\pi$ sector have a single $D$ or $\bar D$ propagator in every loop.
The integral over the loop energy of a $D$ 
can therefore be evaluated by closing the contour around the pole
in the $D$  propagator,
putting the $D$ on its energy shell.
Since $D$ or $\bar D$ lines in loops can be put on their energy shells,
it is sufficient to consider Green functions in which all external 
$D$ or $\bar D$ lines are on their energy shells.

\begin{figure}[tb]
\centerline{\includegraphics*[height=3cm,angle=0,clip=true]{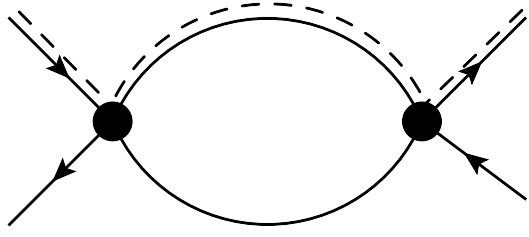}}
\vspace*{0.0cm}
\caption{
The one-loop diagram for the LO $D^{*0} \bar D^0$ transition amplitude.
There is an implied sum over the two possible directions of the arrows on the internal lines.
The diagram with external legs amputated can be expressed as
the contact interaction vertex $(-i C_0/2)\delta^{ij}$
multiplied by $C_0 \Pi(E_{\rm cm})$,
where $\Pi(E_{\rm cm})$ is the LO $D^{*0} \bar D^0$ self-energy.
}
\label{figTwo}
\end{figure}

The amputated connected Green function for $D^{*0} \bar D^0 \to D^{*0} \bar D^0$
in the $C=+$ channel,
with the incoming and outgoing $\bar D^0$ or $D^0$ on their energy shells,
can be expressed as $+i {\cal A}^{ij}$, where the tensor  ${\cal A}^{ij}$
is a function of the total energy $P_0$ and the momenta of the two incoming 
and two outgoing particles.  Its vector indices are associated with the 
polarizations of the incoming and outgoing $D^{*0}$ or $\bar D^{*0}$.
In XEFT, the corresponding amplitude in the $C=-$ channel is 0.
We will refer to ${\cal A}^{ij}$ as
the transition amplitude for $D^{*0} \bar D^0$ in the $C=+$ channel
or simply as the {\it transition amplitude}. 
Its contribution to the amplitude for $D^{*0} \bar D^0 \to D^{*0} \bar D^0$ is
$+i {\cal A}^{ij}/2$, where
the factor of $(1/\sqrt2)^2$ comes from projections onto the $C=+$ channel.

\begin{figure}[tb]
\centerline{\includegraphics*[height=3cm,angle=0,clip=true]{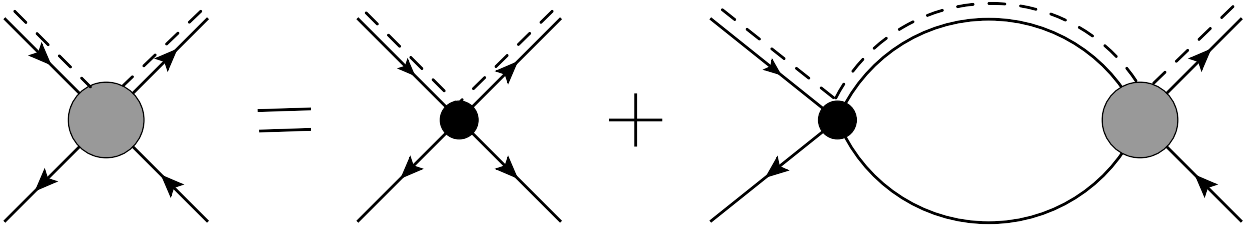}}
\vspace*{0.0cm}
\caption{
The Lippmann-Schwinger integral equation for the LO $D^{*0} \bar D^0$ transition amplitude.
In the second diagram on the right side,
there is an implied sum over the two possible directions of the arrows on the internal lines.
The Feynman rule for the LO transition amplitude is given in Eq.~\eqref{amprule}.
}
\label{fig:LS}
\end{figure}

The tree level term in the LO transition amplitude, which is given by the Feynman rule in 
Eq.~\eqref{vertex:D*D}, is $- C_0 \delta^{ij}$.  
The $D^{*0} \bar D^0$ coupling constant $C_0$ must be treated
nonperturbatively in order to generate the bound state that can be identified 
with the $X(3872)$.  
The one-loop diagram in Fig.~\ref{figTwo} is therefore also LO.
The complete LO transition amplitude can be obtained by summing 
a geometric series of the one-loop diagrams.  
This sum is equivalent to the solution of the 
Lippmann-Schwinger equation shown in Figure~\ref{fig:LS}.

In Galilean-invariant XEFT,
the LO transition amplitude depends on the energies and momenta 
of the initial-state and final-state particles only through the 
Galilean-invariant combination
\beq	
E_{\rm cm} = P_0 - \frac{P^2}{2(2M+m)},
\label{Ecm-D*D}
\eeq
where $P_0$ is the total energy of the $D^{*0}$ and $\bar D^0$ relative 
to the $D^0 \bar D^0 \pi^0$ threshold and $\bm{P}$ is their total momentum.
The LO  transition amplitude  is $(2 \pi/\mu){\cal A}(E_{\rm cm}) \delta^{ij}$, where
\beq
{\cal A}(E_{\rm cm})  = 
\frac{\mu/(2\pi)}{- [C_0^{-1} + \delta C_0^{-1}] + \Pi(E_{\rm cm})}
\label{A0-E:d}
\eeq
and $\Pi(E_{\rm cm})$ is a function of $E_{\rm cm}$ 
that we will refer to as the $D^{*0} \bar D^0$ {\it self-energy}.
The inverse coupling constant from the contact interaction vertex  
in Eq.~\eqref{vertex:D*D} has been separated into $C_0^{-1}$
and a LO counterterm $\delta C_0^{-1}$.
Despite the notation, $\delta C_0^{-1}$ is independent 
of the NLO counterterm $\delta C_0$ in the counterterm vertex  
in Eq.~\eqref{vertex:counterterm}.
With dimensional regularization, the $D^{*0} \bar D^0$ self-energy is 
\beq	
\Pi(E_{\rm cm}) = -  2 \mu
\frac{\Lambda^{3-d} \Gamma(1-d/2)}{(4 \pi)^{d/2}}
\big[ 2 \mu (E_* - E_{\rm cm} )  \big]^{d/2-1} ,
\label{selfD*D}
\eeq
where $\mu$ is the reduced kinetic mass of $D^{*0} \bar D^0$
defined in Eq.~\eqref{mu*}.
If the amplitude ${\cal A}(E_{\rm cm})$ in 
Eq.~\eqref{A0-E:d} has a pole in $P_0$, the residue of the pole is
\beq
Z_X  =
- \frac{(4 \pi)^{d/2-1} \Lambda^{d-3}}{2 \mu \Gamma(2-d/2)} 
\gamma_X^{4-d}.
\label{ZX}
\eeq
This is a smooth function of $d$ for $d<4$.

The $D^{*0} \bar D^0$ self-energy has a linear ultraviolet divergence 
that is manifested by a pole in $d-2$:
\beq
\frac{1}{d-2} \lim_{d \to 2} (d-2) \Pi(E_{\rm cm}) = 
\frac{\mu \Lambda}{\pi (d-2)}.
\label{Pi-PDS}
\eeq
In the minimal power-divergence subtraction (PDS) scheme \cite{Kaplan:1998tg},
this pole is cancelled by a LO counterterm:
\beq
(\delta C_0^{-1})_{\rm PDS} =
\frac{\mu \Lambda}{\pi (d-2)}.
\label{C0-PDS}
\eeq
The LO transition amplitude in Eq.~\eqref{A0-E:d} then has a finite limit as $d \to 2$:
\beq
\lim_{d \to 2} {\cal A}(E_{\rm cm})  = 
\frac{1}{\log \big(2 \mu (E_* - E_{\rm cm})/\Lambda_2^2 \big) \Lambda}.
\label{A0:d->2}
\eeq
The momentum scale $\Lambda_2$ in the logarithm is determined by
the limiting behavior  of $C_0^{-1}$ as $d \to 2$.

In the physical dimension $d = 3$, the LO transition amplitude 
in Eq.~\eqref{A0-E:d} reduces to the form
\beq
\lim_{d \to 3} {\cal A}(E_{\rm cm})  = 
\frac{1}{- \gamma_X + \sqrt{- 2 \mu (E_{\rm cm} - E_*)}},
\label{A0-Ecm}
\eeq
where $\gamma_X$ is a complex constant that we will
refer to as the {\it binding momentum} of the $X(3872)$.
It is determined by
the  limiting behavior of $C_0^{-1}$ as $d \to 3$:
\beq
\gamma_X = (2 \pi/\mu) \lim_{d \to 3}
\left( C_0^{-1} + \delta C_0^{-1} \right).
\label{gamma-C0}
\eeq
The amplitude ${\cal A}(E_{\rm cm})$ in Eq.~\eqref{A0-E:d} 
has a pole in $P_0$ on the physical sheet.
The residue of the pole is
\beq
\lim_{d \to 3} Z_X  =- \gamma_X/\mu.
\label{ZX:d=3}
\eeq

The pole in $P_0$ of the amplitude ${\cal A}(E_{\rm cm})$ 
in Eq.~\eqref{A0-E:d} occurs when $E_{\rm cm}$ equals 
a complex energy $E_X$ that can be interpreted 
as the energy of the $X(3872)$ resonance:
\begin{subequations}
\bea
E_X &=& E_* - \gamma_X^2/(2 \mu) 
\label{EX:complex}\\
&=& \delta - \delta_X - i \Gamma_X/2.
\label{EX:real}
\eea
\label{EX}%
\end{subequations}
In the first line, the complex energy $E_X$ has been expressed in terms 
of the complex $D^{*0}$ energy $E_*$
and the complex binding momentum $\gamma_X$.
In the second line, $E_X$ has been expressed in terms of real variables
$\delta_X$ and $\Gamma_X$ that can be interpreted as the 
binding energy and full width of the $X(3872)$.
The measured value of the binding energy $\delta_X$ 
is given in Eq.~\eqref{EX-exp}. 
The width $\Gamma_X$ has not yet been measured,
but there is an upper bound:
$\Gamma_X < 1.2$~MeV \cite{Choi:2011fc}.
If the complex binding momentum is expressed as
$\gamma_X = \gamma_{\rm re} + i \gamma_{\rm im}$,
where $\gamma_{\rm re}$ and $\gamma_{\rm im}$ are real,
the binding energy and full width are 
\begin{subequations}
\bea
\delta_X  &=& (\gamma_{\rm re}^2 - \gamma_{\rm im}^2)/(2 \mu),
\label{E-X}
\\
\Gamma_X  &=& \Gamma_{*0} +2 \gamma_{\rm re}\gamma_{\rm im}/\mu.
\label{Gamma-X}
\eea
\label{E,Gamma-X}%
\end{subequations}

In calculations beyond LO in the XEFT power counting,
it is convenient to replace the contact interaction vertex in 
Eq.~\eqref{vertex:D*D} by a vertex for the transition amplitude.
The Feynman rules are the same for the 
$D^{*0} \bar D^0 \to D^{*0} \bar D^0$, 
$D ^0 \bar D^{*0} \leftrightarrow D^{*0} \bar D^0$, 
and $D ^0 \bar D^{*0} \to D ^0 \bar D^{*0}$ transition amplitudes:
\beq
\boxed{
+i \left(\frac{1}{\sqrt2} \right)^2 
\frac{2 \pi}{\mu}{\cal A}(E_{\rm cm})\delta^{ij},}
\label{amprule}
\eeq
where $E_{\rm cm}$ is the function of the total energy $P_0$ 
and the total momentum $P$ of the charm meson pair 
defined in Eq.~\eqref{Ecm-D*D} and $i$ and $j$ are the vector indices 
of the incoming and outgoing $D^{*0}$.
In $d$ dimensions, the function ${\cal A}(E_{\rm cm})$
is defined by Eqs.~\eqref{A0-E:d} and \eqref{selfD*D}.
In the physical dimension $d=3$, it reduces to Eq.~\eqref{A0-Ecm}.
The vertex for the transition amplitude
is represented by a blob, as illustrated for the 
$D^{*0} \bar D^0 \to D^{*0} \bar D^0$ channel 
by the Lippmann-Schwinger equation in Fig.~\ref{fig:LS}.

Beyond leading order in the XEFT power counting,
the  transition amplitude ${\cal A}^{ij}(E_{\rm cm},\bm{p}, \bm{p}')$
for $D^{*0} \bar D^0$ in the $C=+$ channel
in Galilean invariant XEFT
is a function of $E_{\rm cm}$ and the relative momenta 
$\bm{p}$ and $\bm{p}'$ of the incoming and outgoing charm mesons.
At NLO, this amplitude has additional ultraviolet divergences.
The renormalizability of XEFT requires that they be cancelled by appropriate choices of the 
constants $\delta C_0$ and $\delta D_0$ in the counterterm vertex 
in Eq.~\eqref{vertex:counterterm}.
A renormalization prescription for the amplitude
${\cal A}^{ij}(E_{\rm cm},\bm{p}, \bm{p}')$ is necessary
to determine the finite parts after the cancellations.
In order to take into account decay modes of the $X(3872)$
with momenta too large to be described explicitly,
it is advantageous to use the
{\it complex on-shell} (COS) renormalization scheme
in which the pole in the total energy $P_0$
is at its physical value.  The complex energy of the pole can be specified 
in terms of the complex binding momentum $\gamma_X$ of the $X(3872)$,
as in Eq.~\eqref{EX:complex}, or in terms of its binding energy $\delta_X$ 
and full width $\Gamma_X$, as in Eq.~\eqref{EX:real}.
The definition of the COS scheme
for the $D^{*0} \bar D^0 \to D^{*0} \bar D^0$ Green function
can be completed by specifying the residue of the pole in $P_0$. 
The amplitude ${\cal A}^{ij}(E_{\rm cm},\bm{p}, \bm{p}')$
has a well-behaved limit as $\bm{p} \to 0$ and $\bm{p}' \to 0$
that is a function of $E_{\rm cm}$ only.
Thus we complete the definition of the COS scheme
by requiring the residue of the pole in $P_0$ 
for the amplitude ${\cal A}^{ij}(E_{\rm cm},0,0)$
to be the same  as for the LO amplitude 
$(2\pi/\mu) {\cal A}(E_{\rm cm}) \delta^{ij}$.
In the physical dimension $d=3$,
the limiting behavior as $P_0$ approaches the pole is 
\beq	
{\cal A}^{ij}(E_{\rm cm},0,0) \longrightarrow
\frac{ (2\pi/\mu)Z_X \delta^{ij}}{[P_0 - P^2/(2(2M+m))] - [E_* - \gamma_X^2/(2 \mu)]},
\label{ampD*D-pole}
\eeq
where $Z_X$ is the LO residue in Eq.~\eqref{ZX:d=3} and
$\gamma_X$ is the complex binding momentum of the $X(3872)$.
The COS scheme can be implemented through specific choices of the 
counterterms  $\delta C_0$ and $\delta D_0$.

In the previous NLO calculations in XEFT
in Ref.~\cite{Fleming:2007rp} and \cite{Jansen:2013cba}, 
the NLO counterterm $\delta C_0$ was included,
but $\delta D_0$ was not.
The counterterm  $\delta C_0$ could have been chosen so that the 
pole in $P_0$ was at the physical point, as in Eq.~\eqref{ampD*D-pole}.
However without the counterterm  $\delta D_0$,
the prescription in Eq.~\eqref{ampD*D-pole} for the residue of the pole 
could not have been implemented. 
In fact, the residue would have been ultraviolet divergent.

\section{$\bm{D^{*0} {\bar D}^0}$ Scattering}
\label{sec:D*Dscat}

In this Section, we use Galilean-invariant XEFT
to calculate the $D^{*0} \bar D^0$ scattering length 
to NLO in the XEFT power counting.

\subsection{NLO Transition Amplitude}
\label{sec:D*DAmp}

In Galilean-invariant XEFT,
the NLO transition amplitude
for  $D^{*0} \bar D^0 \to D^{*0} \bar D^0$ in the $C=+$ channel
in the center-of-momentum frame
is a tensor ${\cal A}^{ij}(E,\bm{p},\bm{p}')$ 
that depends on the total energy $E$ and the relative momenta
$\bm{p}$ and $\bm{p}'$ of the incoming and outgoing charm mesons.
The LO transition amplitude is $(2 \pi/\mu){\cal A}(E) \delta^{ij}$,
where ${\cal A}(E)$ in the physical dimension $d=3$ is given in Eq.~\eqref{A0-Ecm}.
The NLO diagrams for the transition amplitude are calculated in 
Appendix~\ref{sec:NLOAmp}.
The 20 diagrams are labeled $An$, $Bn$, $Cn$, and $Dn$, 
where $n$ is an integer.
The NLO terms in the transition amplitude have a well-behaved limit 
as $\bm{p} \to 0$ and $\bm{p}' \to 0$ that is diagonal in the vector indices $i$ and $j$.
The NLO transition amplitude can be expressed as
\bea
{\cal A}^{ij}(E,0,0) &=& \frac{2 \pi}{\mu} 
\bigg( {\cal A}(E) 
+ (g^2 \mu^2/4m f_\pi^2) \big[ F(E){\cal A}^2(E)  + G(E) {\cal A}(E) \big]
\nonumber \\
&& \hspace{2cm}
+ (C_2/C_0)H(E){\cal A}^2(E)
 - \frac{2\pi [\delta C_0 + \delta D_0 E]}{\mu C_0^2} {\cal A}^2(E)
\bigg) \delta^{ij}.
\label{A-NLO}
\eea
The NLO correction terms either have a factor of ${\cal A}^2(E)$ or a factor of ${\cal A}(E)$.
The function $F(E)$, which has dimensions of momentum, 
is obtained by adding the pion-exchange diagram $A4$ in Eq.~\eqref{A4} and
the  $D^*$ propagator correction diagrams $B1$ and $B2$ in Eqs.~\eqref{B1} and \eqref{B2}.
The function $B(E)$, which is dimensionless, 
is the sum of the pion-exchange diagrams $A2$ and $A3$ in Eq.~\eqref{A23:p=0}. 
The function $H(E)$, which has dimensions of (momentum)$^3$, 
comes from the $\nabla^2$ vertex diagrams $C$ in Eqs.~\eqref{C1234}.
The $\delta C_0$ and $\delta D_0$ terms come from the 
$D^* \bar D$ counterterm diagrams $D$ in Eq.~\eqref{D1234}.
The functions $F(E)$, $G(E)$, and $H(E)$ can be expressed in terms of the 
loop integrals $J_n$, $I_n$, and $K_{lmn}$ defined in Appendix~\ref{sec:LoopInt}:
\begin{subequations}
\bea
F(E) &=&  
\frac{(-8\pi)r}{d}
\Big( \frac{1}{\sqrt{1-r}} 
\big[ 2 K_{110}(E) -2 \mu (2 E_* - r E) K_{111}(E)  
- (2-r) J_1(E)^2 \big]
\nonumber \\ 
&& \hspace{0.5cm}
+ 2 \big[ K_{110}(E) - 2 \mu E_* K_{120}(E) \big]
+ r  I_1(E_*) \big[4 \mu E_* J_2(E) -  d J_1(E)  \big] \Big),
\label{F(E)}
\\
G(E) &=&
\frac{8 r \sqrt{1-r}\, [(E-E_*) J_1(E) -r E I_1(E)]}{d [(1-r)E-E_*]},
\label{G(E)}
 \\
H(E) &=&  
8\pi \mu (E - E_*) J_1(E) ,
\label{H(E)}
\eea
\label{FGH(E)}%
\end{subequations}
where $r = \mu_\pi/\mu$ is the ratio of reduced masses defined in Eq.~\eqref{r-mu}.
The terms proportional to ${\cal A}(E)$ in Eq.~\eqref{A-NLO}
have a single pole in $E$ at $E_* - \gamma_X^2/(2 \mu)$.
The terms proportional to ${\cal A}^2(E)$ have an unphysical double pole in $E$.
The N$^2$LO term would have a triple pole and higher order terms 
would have even higher poles.  These unphysical multiple poles  
can be summed to all orders, in which case they 
produce a shift in the position of the pole in the LO amplitude.
In Ref.~\cite{Jansen:2013cba}, such a resummation was used
in the NLO calculations of the binding energy of the $X(3872)$ 
and its partial width into $D^0 \bar D^0 \pi^0$.
It was not used in the NLO calculation of the $D^{*0} \bar D^0$ scattering length.
An expression for the transition amplitude 
that is accurate to NLO and has only a single pole is
\beq
{\cal A}^{ij}(E,0,0) = 
\frac{(2 \pi/\mu)\big[1 + (g^2 \mu^2/4m f_\pi^2) G(E) \big] \delta^{ij}}
{{\cal A}(E)^{-1} - (g^2 \mu^2/4m f_\pi^2) F(E) 
- (C_2/C_0)H(E)
+(2\pi/\mu C_0^2)[\delta C_0 + \delta D_0 E]}.
\label{A-NLO:frac}
\eeq
To NLO accuracy, the factor of $1 + (g^2 \mu^2/4m f_\pi^2) G$ in the numerator
could equally well be moved to the denominator
as a factor $1 - (g^2 \mu^2/4m f_\pi^2) G$ multiplying
${\cal A}(E)^{-1}$.

\subsection{Renormalization}
\label{sec:Renorm}

Loop integrals in XEFT have ultraviolet (UV) divergences.
With dimensional regularization, the UV divergences 
produce poles in $d-2$ and poles in $d-3$.  
A pole in $d-3$ represents a logarithmic UV divergence,
and a pole in $d-2$ represents a linear UV divergence.
In previous NLO calculations in XEFT \cite{Fleming:2007rp,Jansen:2013cba},
power divergence subtraction was used to remove the poles in $d-2$.
The subsequent limit $d \to 3$ produces terms 
that depend explicitly on the renormalization scale $\Lambda$.  
In the PDS scheme, 
a loop integral with a pole term $\Lambda^{3-d}/(d-2)$
requires a canceling counterterm $-\Lambda/(d-2)$.
In the physical dimension $d= 3$, 
the loop integral becomes independent of $\Lambda$,
but the canceling counterterm reduces to $- \Lambda$.

In the NLO calculation of the decay of the $X(3872)$ into $D^0 \bar D^0 \pi^0$
in Ref.~\cite{Fleming:2007rp}, the explicit dependence on $\Lambda$
was through terms proportional to $\Lambda$, $\Lambda^2$, and $\log \Lambda$.
The explicit dependence on $\Lambda$ was exactly
cancelled by the implicit dependence on $\Lambda$
of the coupling constants of XEFT.
In the NLO calculations of Ref.~\cite{Jansen:2013cba}, 
in addition to terms proportional to $\Lambda$, $\Lambda^2$, $\log \Lambda$,
there was   explicit dependence on $\Lambda$ from  terms proportional to
$1/(\Lambda-\gamma)$, where $\gamma$ is the leading
order binding momentum or inverse scattering length.
These terms  were introduced by a resummation prescription
for dealing with an infrared divergence at the $D^{*0} \bar D^0$ threshold.
The terms proportional to $1/(\Lambda-\gamma)$ were not cancelled
by the implicit dependence on $\Lambda$ of the coupling constants,
but they were suppressed by a factor proportional to 
the partial width $\Gamma_{*0,\pi}$ for $D^{*0} \to D^0 \pi^0$.

The poles in $d-3$ in the NLO  transition amplitude
are given in Appendix~\ref{sec:NLOAmp} in Eqs.~\eqref{A4B1:d->3}.
In the NLO  transition amplitude in Eq.~\eqref{A-NLO},
the only poles in $d-3$ come from the two-loop integrals 
$K_{110}$, $K_{120}$, and $K_{111}$ in the function $F(E)$ in Eq.~\eqref{F(E)}.
The poles in $K_{120}$ and $K_{111}$ are constants 
given in Eqs.~\eqref{K20:E*} and  \eqref{K11:E*}, respectively.
The pole in $K_{110}$ is a linear function of $E$ 
given in Eq.~\eqref{K10:d=3}.
Thus the dependence on the energy in the pole terms is
of the form ${\cal A}^2(E)$ and $E {\cal A}^2(E)$.
These poles in $d-3$ can be cancelled by the 
$\delta C_0$ and $\delta D_0$ counterterms in Eq.~\eqref{A-NLO}.
Thus logarithmic UV divergences  in the NLO  transition amplitude
can be cancelled by the available counterterms.

The poles in $d-2$ in the NLO  transition amplitude
are given in Appendix~\ref{sec:NLOAmp} 
in Eqs.~\eqref{A23C:d->2} and \eqref{A4B:d->2}.
In the NLO  transition amplitude in Eq.~\eqref{A-NLO},
the functions $G(E)$ and $H(E)$ have single poles in $d-2$ 
while the function $F(E)$ has double poles.
The single poles in $d-2$ in the one-loop integrals 
$J_1$ and  $I_1$ are  given in Eqs.~\eqref{J1I1:d->2}.
The poles in $d-2$ in the two-loop integrals 
are given in Eqs.~\eqref{K:d->2}:
$K_{110}$ has a double pole
and $K_{120}$ has a single pole.
The pole terms with the energy dependence 
${\cal A}^2(E)$ and $E {\cal A}^2(E)$ can be cancelled 
by the counterterms $\delta C_0$ and $\delta D_0$, respectively.
They include the double-pole terms from $F(E)$,
some of the single-pole terms from $F(E)$,
and the single-pole terms from $H(E)$.
However there are also single-pole terms from $F(E)$
that have the energy dependence 
$ \log(E_* - E) {\cal A}(E)^2$.
In the single pole of  $G(E)$ in Eq.~\eqref{G(E)},
the energy-dependent denominator is cancelled,
so it gives pole terms with  the energy dependence 
${\cal A}(E)$. Neither of these terms can be cancelled 
by the available counterterms $\delta C_0$ and $\delta D_0$.
The poles in $d-2$ after suitable choices of the counterterms are
\bea	
{\cal A}^{ij}(E) & \longrightarrow&
\left( \frac{g^2}{4m f_\pi^2} \right)
\frac{4 r  \sqrt{1-r}\, \mu \Lambda^2}{d-2}
\log \frac{2 \mu (E_* - E)}{\Lambda^2} 
 {\cal A}^2(E) \, \delta^{ij}
 \nonumber \\
 && + \left( \frac{g^2}{4m f_\pi^2} \right) 
\frac{(-4) r \sqrt{1-r}\,  \mu \Lambda}{d-2}
{\cal A}(E)\,  \delta^{ij}.
\label{A(E):d->2}
\eea
The argument of the logarithm has been made dimensionless 
by using the renormalization scale $\Lambda$.

The uncanceled poles in $d-2$ in 
Eq.~\eqref{A(E):d->2}
seem to indicate that the NLO transition amplitude 
in Galilean-invariant XEFT has linear UV divergences
that cannot be removed by renormalization.
This puzzle can be resolved by taking into account the 
expression for the amplitude ${\cal A}(E)$ in $d=2$, 
which is given in Eq.~\eqref{A0:d->2}.
Its reciprocal ${\cal A}(E)^{-1}$ is linear in $\log(E_* - E)$.
The cancellation between the two terms  in Eq.~\eqref{A(E):d->2}
leaves a term with the energy dependence ${\cal A}^2(E)$
that can be cancelled by the counterterm $\delta C_0$.
 
We proceed to implement the complex on-shell renormalization scheme 
specified by Eq.~\eqref{ampD*D-pole}.
This scheme requires the pole in $E$ of $ {\cal A}^{ij}(E,0,0)$ 
to be at the same complex energy $E_X$ and to have the same residue
as the LO amplitude $(2 \pi/\mu){\cal A}(E) \delta^{ij}$.
It can be implemented as specific choices of the counterterms 
$\delta C_0$ and $\delta D_0$.
We use the variation of the amplitude in Eq.~\eqref{A-NLO:frac}
in which the term $(g^2 \mu^2/4m f_\pi^2) G$ 
in the numerator is moved to the denominator.
The renormalized expression for the transition amplitude is
\bea
{\cal A}^{ij}(E,0,0) 
&=& (2 \pi/\mu) 
\delta^{ij}
\big[
{\cal A}(E)^{-1} 
- (g^2 \mu^2/4m f_\pi^2)  
[ G(E) {\cal A}(E)^{-1} - G(E_X) (E-E_X)/Z_X ]
\nonumber \\
&& \hspace{3.5cm}
- (g^2 \mu^2/4m f_\pi^2)  F_{\rm sub}(E)
- (C_2/C_0) H_{\rm sub}(E) 
\big]^{-1},
\label{A-NLO:p=0}
\eea
where $F_{\rm sub}(E) $ and $H_{\rm sub}(E) $ are obtained by 
subtracting terms from $F(E) $ and $H(E) $:
\begin{subequations}
\bea
F_{\rm sub}(E) 
&=& F(E) - F(E_X) - F'(E_X) (E-E_X),
\label{Fsub}
\\
H_{\rm sub}(E) &=& H(E) - H(E_X) - H'(E_X) (E-E_X).
\label{Hsub}
\eea
\label{FHsub}%
\end{subequations}
The denominator in Eq.~\eqref{A-NLO:p=0} vanishes at $E=E_X$.
The value of its first derivative at $E=E_X$ is such that the residue of the 
pole in $E$ is $(2 \pi/\mu)Z_X \delta^{ij}$, 
in accord with the renormalization prescription in Eq.~\eqref{ampD*D-pole}.
The explicit expressions for the counterterms are given by
\begin{subequations}
\bea
\frac{2\pi \delta C_0}{\mu C_0^2}  &=& 
\frac{g^2 \mu^2}{4m f_\pi^2}
\Big( F(E_X) - \big[ F'(E_X)+  G(E_X) /Z_X \big] E_X \Big)
\nonumber \\
&& \hspace{2cm}
+ \frac{C_2}{C_0} 
\Big( H(E_X) - H'(E_X)  E_X \Big),
\label{dC0}
\\
\frac{2\pi \delta D_0}{\mu C_0^2}  &=& 
\frac{g^2 \mu^2}{4m f_\pi^2} 
 \big[ F'(E_X)+ G(E_X)/Z_X \big]
+  \frac{C_2}{C_0} H'(E_X).
\label{dD0}
\eea
\label{dC0D0}%
\end{subequations}

In the previous NLO calculations in XEFT
in Refs.~\cite{Fleming:2007rp} and \cite{Jansen:2013cba}, 
power divergence subtraction was used to make 
double and single poles in $d-2$ explicit as terms proportional to 
$\Lambda^2$ and $\Lambda$, where $\Lambda$ is the renormalization scale.
Since the poles in $d-2$ have the same energy dependence  as the 
counterterms  $\delta C_0$ and $\delta D_0$,
the terms proportional to $\Lambda^2$ and $\Lambda$ are
exactly cancelled by those counterterms in the COS scheme.
Thus it is  unnecessary to use power divergence subtraction 
to make the poles in $d-2$ explicit.

\subsection{$\bm{D^{*0} {\bar D}^0}$ Scattering Length}
\label{sec:D*Das+}

T-matrix elements in XEFT involving charm mesons and $\pi^0$'s
can be obtained from the amputated connected Green functions
by the following steps:
\begin{itemize}
\item
Multiply by a polarization vector $\varepsilon^i$
for a $D^{*0}$ or $\bar D^{*0}$ in the initial state
and by a polarization vector ${\varepsilon'}{}^j{}^*$
for a $D^{*0}$ in the final state.
\item
Multiply by a residue factor $|Z_*|^{1/2}$ 
for each $D^{*0}$ or $\bar D^{*0}$ in the initial and final states.
In the COS scheme, $Z_* = 1$.
\item
Put the external lines on their energy shells.
For the $D^0$ or $\bar D^0$ and for the $\pi^0$, the energy shells are real: 
$p_0 = p^2/(2M)$ and $p_0 = p^2/(2m)$, respectively.
For the $D^{*0}$ or $\bar D^{*0}$, the energy shell is complex: 
$p_0 = E_* + p^2/(2(M+m))$.
\end{itemize}

We consider the elastic scattering of $D^{*0} \bar D^0$
in the center-of-momentum frame
with incoming momenta $\pm p \bm{\hat z}$
and outgoing momenta $\pm p \bm{\hat n}$.
The scattering angle satisfies $\cos \theta = \bm{\hat z} \cdot \bm{\hat n}$.
The initial and final polarization vectors of the $D^{*0}$ 
are $\bm{\varepsilon}$ and $\bm{\varepsilon}'$.
The total energy is $E_p = E_* + p^2/(2 \mu)$.
The T-matrix element is 
\beq
{\cal T}(p, \theta) =  
(1/2)\sum_{ij} {\cal A}^{ij}(E_p,p\bm{\hat z},p\bm{\hat n}) 
 \varepsilon^i  {\varepsilon'}{}^j{}^* .
\label{T-NLO}
\eeq
In the limit of zero relative momentum, the scattering is isotropic.
The T-matrix element reduces to
\beq
{\cal T}(p=0) =  (1/2) \sum_{ij} {\cal A}^{ij}(E_*,0,0) 
 \varepsilon^i  {\varepsilon'}{}^j{}^* ,
\label{T-NLO:p=0}
\eeq
where the transition amplitude is given by
Eq.~\eqref{A-NLO:p=0} evaluated at $E=E_*$.

In the case of short-range interactions,
a scattering amplitude can be expanded in powers of 
 the relative momentum.
This expansion is called the {\it effective range expansion}.
Beyond LO in XEFT, the effective range expansion
for $D^{*0} \bar D^0$ breaks down because of the effects 
of the exchange of a pion that can be on its energy shell.
However Jansen, Hammer, and Jia pointed out that the 
leading term in the effective range expansion,
which is the S-wave $D^{*0} \bar D^0$ scattering length, 
remains well defined \cite{Jansen:2013cba}.
They calculated the scattering length to NLO in original XEFT,
truncating the expression at first order in an expansion 
in powers of $\gamma_X/\sqrt{2m \delta}$
and at leading order in  $m/M$.
We will calculate the scattering length to NLO in Galilean-invariant XEFT,
truncating the expression at fourth order in an expansion 
in powers of $\gamma_X/\sqrt{2 \mu E_*}$
but keeping all orders in $m/M$.

The S-wave scattering length $a_{s,+}$ for $D^{*0} \bar D^0$ in the $C=+$ channel 
can be defined by expressing the T-matrix in the zero-momentum limit 
in Eq.~\eqref{T-NLO:p=0} as 
\beq
{\cal T}(p=0) = 
(1/2) (2  \pi / \mu)(- a_{s,+})
 \bm{\varepsilon} \cdot \bm{\varepsilon}'{}^*.
\label{T-NLO:as+}
\eeq
At leading order in XEFT, the inverse scattering length is equal to the 
complex binding momentum:  $1/a_{s,+} = \gamma_X$.
The inverse scattering length at NLO is
\bea
1/a_{s,+} &=& 
\gamma_X
+  (g^2 \mu^2/4m f_\pi^2)\big[ G(E_X) - 2 G(E_*)  \big] \gamma_X/2 
\nonumber \\
&& \hspace{1cm}
+(g^2 \mu^2/4m f_\pi^2)  F_{\rm sub}(E_*) 
+ (C_2/C_0) H_{\rm sub}(E_*).
\label{as+}
\eea
We have used the expressions for $E_X$ and $Z_X$ 
in terms of $\gamma_X$ in Eqs.~\eqref{EX:complex} and \eqref{ZX:d=3}.
This expression for  $1/a_{s,+}$ in Eq.~\eqref{as+}
can be simplified by expanding it in powers of $\gamma_X/\sqrt{2 \mu E_*}$
using the expressions for the one-loop integrals $J_n$ and $I_n$
in Eqs.~\eqref{J1:d=3} and \eqref{I1:d=3} and the threshold expansions for the 
two-loop integrals $K_{lmn}$ in Eqs.~\eqref{K:thresh}.
The expansion of $G(E_X)- 2 G(E_*)$
to third order in  $\gamma_X$ is
\beq
G(E_X)- 2 G(E_*) \approx 
- \frac{2i}{3 \pi}  r^{3/2} \sqrt{1-r}\, \sqrt{2 \mu E_*}
\left( 1 + \frac{(2+r) \gamma_X ^2}{2r (2 \mu E_*)}
- \frac{i \gamma_X ^3}{r^{3/2} (2 \mu E_*)^{3/2}}  \right),
\label{GsubEX}
\eeq
where $r= \mu_\pi/\mu$ is the ratio of reduced masses defined in Eq.~\eqref{r-mu}.
The quantity $H_{\rm sub}(E_*)$ is very simple:
\beq
H_{\rm sub}(E_*) = \gamma_X^3/2.
\label{HsubE*}
\eeq
The expansion of $F_{\rm sub}(E_*) $ to fourth order  in $\gamma_X$ is
\bea
F_{\rm sub}(E_*)  &\approx& 
-i \frac{r^{5/2} [1- \sqrt{1-r}\,]}{6 \pi \sqrt{1-r}}  \gamma_X \sqrt{2 \mu E_*}
\nonumber \\
&& + \frac{1}{12 \pi^2} 
\left( - \frac{r^{5/2}}{5} + \frac{(12-22r-5r^2)r^{1/2}}{3\sqrt{1-r}} 
+ \frac{(2-r)(2-4r-r^2) \arccos(\sqrt{r}\,)}{1-r}\right.
\nonumber
\\
&& \hspace{2cm} \left.
+ \frac{8(2-r)}{\sqrt{1-r}}
 {}_2F_1\big(\mbox{$-\frac12$},\mbox{$-\frac12$},\mbox{$\frac32$},1-r\big)
 \right) \frac{\gamma_X ^4}{2 \mu E_*}.
\label{FsubE*}
\eea
Our final result for the inverse scattering length,
including all terms suppressed by up to three powers of 
$\gamma_X/\sqrt{2 \mu E_*}$, is
\bea
1/a_{s,+} &=& \gamma_X
+ \frac{C_2}{2C_0}\gamma_X^3
+ \frac{g^2 \mu^2}{12 \pi m f_\pi^2} 
\Bigg[ 
- i r^{3/2} \left(1 - \frac12 r + \frac18 r^2 \right) \gamma_X \sqrt{2  \mu E_*}
\nonumber \\
&&\hspace{4.5cm}
- i r^{1/2}\left(1 - \frac38 r^2 \right) 
\frac{\gamma_X^3}{\sqrt{2  \mu  E_*}}
\nonumber \\
&&\hspace{4.5cm}
+ \left(1 - \frac12 r - \frac18 r^2 - \frac{1}{10\pi} r^{5/2} \right)  
\frac{\gamma_X^4}{2 \mu E_*} \Bigg] .
\label{as+:gammaX}
\eea
The coefficient of each term has been expanded to relative order $r^{5/2}$.
The leading power of $r$ in the $g^2  \gamma_X$ term
comes from the function $G$.
The entire $g^2\gamma_X^3$ term comes from the function $G$.
The $g^2 \gamma_X^4$ term receives contributions of order
$r^0$ from both the function $F$ and the function $G$.

\subsection{Comparison with Previous Calculation}
\label{sec:compare}

We can compare our result for the scattering length $a_{s,+}$
in Eq.~\eqref{as+:gammaX} with that from a previous calculation by 
Jansen, Hammer, and Jia \cite{Jansen:2013cba}.
They used original XEFT to calculate the complex energy $E_X$ of the $X(3872)$
and the complex scattering length $a_{s,+}$,
dropping terms that were suppressed by at least one power of $r = m/M$.
They expressed their results in terms of a
leading-order inverse scattering length $\gamma$.
Their physical renormalization condition was that the
real part of the complex energy  $E_X$ should be identified with the
energy of the $X(3872)$ resonance.
Their NLO corrections included terms proportional to 
$g^2$, $C_2$, and an additional coupling constant $D_2$
that takes into account the dependence of the contact interaction on the pion mass.
The coupling constant $D_2$ allowed them to study the dependence 
of the mass of the $X(3872)$ on the light quark masses.
Their NLO corrections also included terms involving
the partial width $\Gamma_{*0,\pi}$ for $D^{*0} \to D^0 \pi^0$
that were resummed to all orders.
If the other NLO corrections are omitted, their results reduce to
\begin{subequations}
\bea
E_X  &=& 
\delta - \gamma^2/(2 \mu) - i \Gamma_{*0,\pi}/2,
\label{EX-JHY:LO}
\\
a_{s,+}  &=& 1/\left(\gamma -\sqrt{-i \mu \Gamma_{*0,\pi}} \right),
\label{as-JHY:LO}
\eea
\label{JHY:LO}%
\end{subequations}
where $\Gamma_{*0,\pi}$ is the partial width of $D^{*0}$ into  $D^0 \pi^0$.
The universal relation expressing the 
binding energy in terms of a large positive scattering length $a_{s,+}$ is
\beq
E_X = \delta  - 1/(2 \mu a_{s,+}^2) .
\label{EX-universal}
\eeq
The leading order results in Eqs.~\eqref{JHY:LO}
are incompatible with the analytic continuation
of this universal relation to a complex scattering length $a_{s,+}$.
This incompatibility suggests that the resummation 
used in Ref.~\cite{Jansen:2013cba}
to eliminate an infrared divergence at the $D^{*0} \bar D^0$
threshold was inadequate.
Because of this problem,
we will only compare results in the limit $\Gamma_{*0,\pi} \to 0$.
In this limit, their NLO results reduce to
\begin{subequations}
\bea
E_X  &=& 
\delta - \frac{\gamma^2}{2 \mu} \left( 1 +
\frac{(C_2/C_0^2)(4 \pi \gamma/\mu)+(g^2 \mu/6 \pi f_\pi^2) (1/\gamma)
[ (\Lambda- \gamma)^2  - 2 m \delta d_2' ]}
{1 + (g^2 \mu/6 \pi f_\pi^2)(m \delta/\gamma)\log(1 + 2 i \gamma/\sqrt{2 m \delta})} 
\right),
\label{EX-JHY:NLO}
\\
a_{s,+}  &=& \frac{1}{\gamma} 
- \frac{1}{\gamma^2} \left( \frac{g^2 \mu}{12 \pi f_\pi^2} \right)  
\left[ - 2 i \gamma \sqrt{2 m \delta}+ \Lambda^2 - 2 \gamma \Lambda- 2 m \delta d_2' \right],
\label{as-JHY:NLO}
\eea
\label{JHY:NLO}%
\end{subequations}
where $\Lambda$ is the PDS renormalization scale and $d_2'$
is a finite constant related to the coupling constant $D_2$.
Note that the dependence on $\Lambda$ can be eliminated
from both $E_X$ and from $a_{s,+}$ by absorbing it into $d_2'$.
ln the expression for $a_{s,+}$ in Eq.~\eqref{as-JHY:NLO},
the factor of $1/\gamma^2$ comes from a term that has an 
unphysical double pole in the energy.
The double pole could have been eliminated 
in favor of a shift in the position of the single pole by expressing $a_{s,+}$
as the ratio of a numerator and denominator with  NLO accuracy,
as in their result for $E_X$ in Eq.~\eqref{EX-JHY:NLO}.

The complex on-shell renormalization scheme can be implemented by 
choosing $d_2'$ so that the numerator of the NLO correction term 
to $E_X$ in Eq.~\eqref{EX-JHY:NLO} is zero.
The complex energy $E_X$ then reduces to $\delta - \gamma^2/(2 \mu)$,
so $\gamma$ can be identified with the binding momentum $\gamma_X$
of the $X(3872)$.
The NLO scattering length calculated in Ref.~\cite{Jansen:2013cba} reduces to
\beq
a_{s,+}  =
 \frac{1}{\gamma_X} \left[ 1 
+  \frac{2 \pi C_2}{\mu C_0^2} \gamma_X
+ \frac{g^2 \mu}{12 \pi f_\pi^2}
\left(  2 i \sqrt{2 m \delta}  - \gamma_X \right)
\right].
\label{as:JHY}
\eeq
This can be compared with the result from Galilean-invariant XEFT 
in Eq.~\eqref{as+:gammaX}.
Dropping terms suppressed by three or more powers of $\gamma_X/\sqrt{2 \mu E_*}$ 
and keeping only the leading power of $r$ in the coefficients,
our result reduces to
\beq
a_{s,+} = \frac{1}{\gamma_X} 
\left[ 1 - \frac{C_2}{2C_0} \gamma_X^2
+ \frac{g^2 \mu^2}{12 \pi m f_\pi^2} 
\left( i r^{3/2}\sqrt{2  \mu E_*}  +  i r^{1/2} \frac{\gamma_X^2}{\sqrt{2  \mu E_*}}
\right) \right].
\label{as+:simple}
\eeq
If we ignore the tiny difference between $E_*$ and $\delta$
and if we approximate $\mu$ by $M/2$ and $\mu_\pi$ by $m$,
the $\sqrt{2  \mu E_*}$ term in Eq.~\eqref{as+:simple}
agrees with the $ \sqrt{2 m \delta}$ term in Eq.~\eqref{as:JHY}.
However the other correction terms in Eq.~\eqref{as:JHY}
are proportional to $\gamma_X$, while those in
Eq.~\eqref{as+:simple} are proportional to $\gamma_X^2$.
Furthermore the coupling constants $C_2$ and $C_0$
appear in the combination $C_2/C_0^2$ in  Eq.~\eqref{as:JHY},
while they appear in the combination $C_2/C_0^2$ 
in Eq.~\eqref{as+:simple}.
Thus there seems to be a clear discrepancy between the results 
for $a_{s,+}$ in Eqs.~\eqref{as:JHY} and \eqref{as+:simple}.
However the conclusion that there is a discrepancy 
relies on an implicit assumption that the coupling constant
$C_2$  in Eqs.~\eqref{as:JHY} and \eqref{as+:simple} can be identified.
In the following section, we will show the surprising result 
that in the limit $g\to 0$ in which the pions decouple, the 
NLO correction terms proportional to $C_2$
in Eqs.~\eqref{as:JHY} and \eqref{as+:simple}
are indeed compatible.  Whether the apparent discrepancy 
between the  NLO correction terms proportional to $g^2 \gamma_X$
in Eq.~\eqref{as:JHY} and to $g^2 \gamma_X^2$ in Eq.~\eqref{as+:simple} is real
cannot be determined definitively using the results given in Ref.~\cite{Jansen:2013cba}.

\subsection{Decoupled Pions}
\label{sec:pidecouple}

If pions are decoupled by setting $g=0$,
the NLO terms in the transition amplitude  ${\cal A}^{ij}(E,\bm{p},\bm{p}')$ 
reduces to the  $\nabla^2$ vertex diagrams $C$  in Eqs.~\eqref{C1234}
and the $D^* \bar D$ counterterm diagrams $D$ in Eq.~\eqref{D1234}.
The terms proportional  to ${\cal A}^2(E)$
which have a double pole in $E$, can to NLO accuracy be
replaced by terms in the denominator, as in Eq.~\eqref{A-NLO:frac}.
After implementing the COS renormalization scheme,
the transition amplitude reduces to
\beq
{\cal A}^{ij}(E,p,p') = 
\frac{ (2 \pi/\mu) \big[1 + \frac12 (C_2/C_0) \big(p^2 + {p'}^2 \big) \big]}
{{\cal A}(E)^{-1} - (C_2/C_0) H_{\rm sub}(E)}  \delta^{ij},
\label{A-nopi:frac}
\eeq
where $H_{\rm sub}(E)$ is defined by Eqs.~\eqref{Hsub} and \eqref{H(E)}.

To obtain the T-matrix element in Eq.~\eqref{T-NLO},
we put the charm mesons on their energy shells by setting $p'=p$
and setting the energy equal to $E_p=E_*+p^2/(2\mu)$.  
The term $H_{\rm sub}(E_p)$ in the denominator of Eq.~\eqref{A-nopi:frac} is
\beq
H_{\rm sub}(E_p) = 
 \mbox{$\frac12$} \gamma_X^3 + \mbox{$\frac32$} \gamma_X  p^2 + i p^3.
\label{HsubEp}
\eeq
The T-matrix element reduces to
\beq
{\cal T}(p) =  
\frac{ ( \pi/\mu) \big[1 + (C_2/C_0) p^2 \big]}
{- \gamma_X - i p - (C_2/C_0) 
[ \mbox{$\frac12$} \gamma_X^3 +\mbox{$\frac32$} \gamma_X  p^2 + i p^3] }  
\bm{\varepsilon} \cdot \bm{\varepsilon}'{}^*.
\label{T-nopi}
\eeq
To NLO accuracy, the factor $1 + (C_2/C_0) p^2$ in the numerator
of Eq.~\eqref{T-nopi} could equally be included as a factor
$1 - (C_2/C_0) p^2$ multiplying $- \gamma_X - i p$ in the denominator:
\beq
{\cal T}(p) =  
\frac{ \pi/\mu}
{- [\gamma_X + \mbox{$\frac12$}(C_2/C_0) \gamma_X^3 ] 
- \mbox{$\frac12$} (C_2/C_0) \gamma_X p^2 - i p}  
\bm{\varepsilon} \cdot \bm{\varepsilon}'{}^*.
\label{T-nopi:2}
\eeq
We can read off the S-wave inverse scattering length and effective range 
in the $C=+$ channel as coefficients in the denominator:
\begin{subequations}
\bea
1/a_{s,+}  &=& 
\gamma_X + \mbox{$\frac12$}(C_2/C_0) \gamma_X^3,
\label{as+:nopi}
\\
r_{s,+}  &=&  - (C_2/C_0) \gamma_X.
\label{rs+:nopi}
\eea
\label{ars+:nopi}%
\end{subequations}
The inverse scattering length in Eq.~\eqref{as+:nopi} 
is consistent to NLO accuracy with setting $g=0$ in Eq.~\eqref{as+:gammaX}.
Eliminating $C_2/C_0$, we reproduce a familiar relation 
between the scattering length, effective range, and binding momentum
to first order in the range expansion:
\beq
1/a_{s,+}  = 
\gamma_X \left(1 - \mbox{$\frac12$} r_{s,+} \gamma_X\right).
\label{as-nopi}
\eeq

Our result for the effective range  in Eq.~\eqref{rs+:nopi}
implies that, in the absence of pions, the ratio $C_2/C_0$ of coupling constants
has a simple physical interpretation in terms of the effective range
and the binding momentum:
\beq
C_2/C_0  = - r_{s,+}/\gamma_X.
\label{Cratio-nopi}
\eeq
In the previous NLO calculations in XEFT
in Refs.~\cite{Fleming:2007rp} and \cite{Jansen:2013cba},
power divergence subtraction was used 
and $C_0$ and $C_2$ were determined as functions
of the renormalization scale $\Lambda$.
The dependence on $\Lambda$ canceled in the following combination 
of coupling constants:
\beq
C_2/C_0^2  = \mu r_{s,+} /(4 \pi).
\label{Cratio-old}
\eeq
The apparent incompatible with our result in Eq.~\eqref{Cratio-nopi}
is at first puzzling. Our coupling constants
$C_0$ and $C_2$ can be defined as the coefficients of the terms 
in the Lagrangian given in Eq.~\eqref{LD*D}.
This seems to be equivalent to the definitions of $C_0$ and $C_2$
in Ref.~\cite{Fleming:2007rp}
up to corrections suppressed by powers of  $r=\mu_\pi/M_\pi$.
The resolution is that the absence of the counterterm $\delta D_0$
in the Lagrangian for original XEFT in Ref.~\cite{Fleming:2007rp} 
implies that the definitions are not equivalent.
To determine the relation between the coupling constants, 
it is necessary to compare physical observables, such as the effective range $r_{s,+}$.
By equating  $r_{s,+}$ in Eqs.~\eqref{Cratio-nopi} and \eqref{Cratio-old},
we can infer that the
combination $C_2/C_0^2$ in Ref.~\cite{Fleming:2007rp} 
should actually be identified with our ratio $C_2/C_0$ multiplied by
$- \mu \gamma_X/(4 \pi)$.
This identification brings the NLO correction terms to $a_{s,+} $ 
proportional to $(C_2/C_0^2) \gamma_X$ in Eq.~\eqref{as:JHY} 
and $(C_2/C_0) \gamma_X^2$ in Eq.~\eqref{as+:simple}
into agreement.

\section{Summary}
\label{sec:summary}

We have introduced a  Galilean-invariant formulation of XEFT to exploit
the fact that mass is very nearly conserved in the transition $D^* \to D \pi$.
Galilean invariance requires the kinetic mass of the $D^*$ 
to be equal to the sum of the kinetic masses $M$ and $m$ of the $D$ and $\pi$.
One advantage of Galilean invariance is that an amplitude is
the same in every Galilean frame.
A more important advantage is that it strongly constrains the form of ultraviolet divergences.
The operators in the NLO Lagrangian for Galilean-invariant XEFT
can be obtained by making minor modifications 
in the terms of the NLO Lagrangian for original XEFT.
NLO calculations in Galilean-invariant XEFT are ultraviolet finite
for arbitrary values of the masses $M$ and $m$.
In contrast, NLO calculations in original XEFT are ultraviolet finite
only if they are expanded in powers of $m/M$ and truncated at a sufficiently low order.

We also introduced the complex on-shell (COS) renormalization prescription that 
takes into account the effects of transitions to states that cannot be described 
explicitly in XEFT.  These transitions include the decay $D^{*0} \to D^0 \pi^0$,
which accounts for about 38\% of the total width of the $D^{*0}$.
They also include the decays of $X(3872)$  to all final states
other than $D^0 \bar D^0 \pi^0$. These other decay modes
may account for a substantial fraction of the total width of the $X(3872)$.
The COS prescription requires as input the total width of the $D^{*0}$
and the total width of the $X(3872)$, which has not yet been measured.

We illustrated the use of Galilean-invariant XEFT by calculating 
the transition amplitude for $D^{*0} \bar D^0 \to D^{*0} \bar D^0$ to NLO
in the XEFT power counting. We used dimensional regularization
in $d$ spatial dimensions to regularize the ultraviolet divergences.
The NLO Lagrangian includes counterterms that can cancel terms 
whose energy dependence has the form ${\cal A}^2(E)$
and $E {\cal A}^2(E)$, where ${\cal A}(E)$ is the LO transition amplitude.
It was straightforward to verify that the
logarithmic ultraviolet divergences, which appeared as poles in $d-3$,
could be cancelled by the counterterms.
The cancellation of linear ultraviolet divergences, 
which appeared as poles in $d-2$, was not as straightforward.
In addition to the poles that could be cancelled by the counterterms,
there were additional poles in $d-2$
whose energy dependence had the form ${\cal A}^2(E) \log(E_*-E)$
and ${\cal A}(E)$.  They cancelled each other only upon using the 
specific form of the LO transition amplitude ${\cal A}(E)$  for  $d=2$.

We used our NLO result for the transition amplitude to calculate the 
$D^{*0} \bar D^0$ scattering length $a_{s,+}$ to NLO.
Our result differs in several respects from a previous result for $a_{s,+}$
calculated using original XEFT in Ref.~\cite{Jansen:2013cba}.
One important difference is in the effect of the $D^{*0}$ width.  
Our COS renormalization prescription 
ensures that the effect of the $D^{*0}$ width is 
in accord with the universal relation between the
binding energy of an S-wave bound state near threshold
and the scattering length of its constituents.
The effect of the partial width of the $D^{*0}$ into $D^0 \pi^0$  
in Ref.~\cite{Jansen:2013cba} is incompatible with the universal relation.
This suggests that the resummation used in Ref.~\cite{Jansen:2013cba}
to deal with an infrared divergence at the $D^{*0} \bar D^0$
threshold was inadequate.
There is an apparent disagreement between our result and that of 
Ref.~\cite{Jansen:2013cba} in the  NLO corrections 
to $a_{s,+}$ proportional to $C_2$, but we showed that 
the difference is actually due to different definitions of $C_2$.
Our NLO correction to $a_{s,+}$ proportional to $g^2$ was expanded 
in powers of $\gamma_X/ \sqrt{2m\delta}$, 
where $\gamma_X$ is the binding momentum of the $X(3872)$. 
The leading term is proportional to $g^2 \sqrt{2m\delta}$,
and it agrees with that in Ref.~\cite{Jansen:2013cba} in the limit $m/M \to 0$.
The next term in our expansion is second order in $\gamma_X$,
while there is a term in Ref.~\cite{Jansen:2013cba}
that is  first order in $\gamma_X$. 
The origin of this discrepancy has not been determined.

Our Galilean-invariant formulation of XEFT removes obstacles 
to precise calculations of the properties of the $X(3872)$ resonance.
Our COS renormalization prescription takes into account
 the 38\% branching fraction 
of $D^{*0}$ into $D^0 \pi^0$  and the significant branching fraction 
of $X(3872)$ into decay modes other than $D^0 \bar D^0 \pi^0$.
In  original XEFT, there was a serious limitation on the precision
from the need to truncate the expansion
in $m/M$ in order to avoid ultraviolet divergences.
Our Galilean-invariant formulation makes this expansion unnecessary.
The expansion may still be useful to obtain simple results,
but it can be carried out to whatever order is required for the desired precision. 

XEFT was invented to allow the effects of pion transitions 
on the $X(3872)$ resonance to be taken into account systematically.
XEFT can also be used as a framework for 
quantifying the effects of physics beyond the $D \bar D \pi$ sector.
This physics include charmonium states such as the $\chi_{c1}(2P)$,
which couples to $D^* \bar D$, and $\psi(3770)$, which  couples to $D \bar D$.
It also includes tetraquark $c \bar c$ mesons and hybrid $c \bar c$ mesons.
The improved precision  of Galilean-invariant XEFT 
increases the motivation for using this effective field theory
to quantify the effects of physics beyond the $D \bar D \pi$ sector.

\begin{acknowledgments}
This research was supported in part by the Department of Energy 
under grant DE-SC0011726 and by the Simons Foundation.
I thank Hans-Werner Hammer for useful comments.
\end{acknowledgments}

\begin{appendix}

\section{Loop Integrals}
\label{sec:LoopInt}

In this Appendix, we evaluate the loop integrals that are required
to calculate the NLO transition amplitude using dimensional regularization.
We also determine their poles in $d-2$ and $d-3$,
where $d$ is the number of spatial dimensions.

\subsection{Reduction to momentum integrals}

The loop integrals have $D^*$, $D$, and $\pi$ propagators.
Every loop has a $D$ propagator, and it  is convenient to take 
its energy and momentum to be the loop integration variables.
A $D$ propagator has a denominator of the form 
$p_0 - p^2/(2M) + i \epsilon$.  The integral over $p_0$
can be evaluated by closing the contour around the pole at $p_0 = p^2/(2M)$,
which puts the $D$ on its energy shell.
With dimensional regularization in $d$ spatial dimensions,
the measure for a momentum integral is
\beq
\int_{\bm{p}} \equiv \Lambda^{3-d} \int \frac{{\rm d}^dp}{(2 \pi)^d},
\label{int-p}
\eeq
where $\Lambda$ is the renormalization scale.
Tensor reduction can be used to reduce the loop momentum integrals
to scalar integrals. Whenever possible, it is best to express the 
momentum-dependent terms in the numerator
as a linear combination of the propagator denominators.
After canceling terms in the numerator with propagators,
the loop integrand reduces to a product of propagators.

The propagators  in the loop integrand depend on the complex $D^{*0}$ energy
$E_*$ and on the masses $m$ and $M$ of the $\pi^0$ and $D^0$.
The loop integrals depend on the masses through the 
reduced $D^* \bar D$ kinetic mass $\mu$ defined in Eq.~\eqref{mu*} 
and the reduced $D \pi$ mass $\mu_\pi$ defined in Eq.~\eqref{mupi}.
It is convenient to express the dependence on the masses in terms of 
$\mu$ and the ratio $r = \mu_\pi/\mu$ defined in Eq.~\eqref{r-mu}.

In loop integrals that have cuts with $D \bar D \pi$ intermediate states,
after integrating over the $D$ and $\bar D$ energies,
the pion propagator depends on the momenta $\bm{p}$ and $\bm{q}$
of the $D$ and $\bar D$ and on the masses $m$ and $M$.  
In evaluating the loop integral, it is often convenient
to express the pion propagator in a form that depends on reduced masses 
instead of masses:
\bea
&& \frac{i}{E - (\bm{p} + \bm{q})^2/(2m) - (p^2 +q^2)/(2M) + i \epsilon}
\nonumber
\\
&& \hspace{1cm}
= \frac{i}{E - (\bm{q} + \sqrt{1-r}\, \bm{p})^2/(2\mu_\pi) - p^2/(2 \mu)  + i \epsilon}.
\label{piprop}
\eea
Unlike the first expression for the pion propagator in Eq.~\eqref{piprop},
the second expression is not manifestly invariant
under the interchange $\bm{p} \leftrightarrow \bm{q}$.

\subsection{One-loop momentum integrals}
\label{sec:1loop}

The one-loop momentum integrals whose integrands are a $D^*$ propagator
raised to an integer power have the form
\beq
J_n(E) = \int_{\bm{p}} \frac{1}{[ p^2 - 2\mu (E - E_*)]^n} .
\label{int-Jn}
\eeq
The analytic result for this integral is
\beq
J_n(E)  = \frac{\Lambda^{3-d} \Gamma(n-d/2)}{(4 \pi)^{d/2}} 
[ 2 \mu (E_* -E) ]^{d/2-n}.
\label{Jn}
\eeq
The branch cut in $E$ is determined by the 
negative imaginary part of the complex energy $E_*$.

The one-loop momentum integrals whose integrands are a $\pi$ propagator
raised to an integer power have the form
\beq
I_n(E)=\int_{\bm{p}} \frac{1}{[p^2 - 2\mu_\pi E - i \epsilon]^n}.
\label{int-In}
\eeq
The analytic result for this integral is
\beq
I_n(E)  = r^{d/2-n}
 \frac{\Lambda^{3-d} \Gamma(n-d/2)}{(4 \pi)^{d/2}} 
\left[ e^{-i \pi} \,  2 \mu E \right]^{d/2-n}.
\label{In}
\eeq
where $r= \mu_\pi/\mu$ is the ratio of the reduced masses 
defined in Eq.~\eqref{r-mu}.
The branch cut in $E$ is specified by the $i \epsilon$ prescription
in Eq.~\eqref{int-In}.
Since we choose the zero of energy to be at the $D^0 \bar D^0 \pi^0$ threshold,
we need only consider positive values for the real part of $E$.

The one-loop momentum integrals whose integrands have a single $\pi$ propagator
and a $D^*$ propagator raised to an integer power are
\beq
L_n(E,p) = \int_{\bm{q}} 
\frac{1}{[q^2 - 2\mu (E - E_*)]^n}  \, 
\frac{(2 \mu)^{-1}}{ (\bm{p} + \bm{q})^2/(2m) + (p^2 + q^2)/(2M) - E -  i \epsilon} .
\label{int-Ln}
\eeq
The function $L_0$ can be expressed in terms of the integral $I_1$ 
given by Eq.~\eqref{In}:
\beq
L_0(E,p) = r \, I_1(E-p^2/2 \mu).
\label{L0}
\eeq
This result can be obtained most easily by using the expression 
for the pion propagator on the right side of Eq.~\eqref{piprop},
and then shifting the integration variable $\bm{q}$.
The function $L_1$ can be expressed as a Feynman parameter integral:
\bea
L_1(E,p) &=& r^{-d/2} 
\frac{\Lambda^{3-d} \Gamma(2-d/2)}{(4 \pi)^{d/2}} 
\int_0^1 {\rm d}x\,   \big( x+(1-x)r \big)^2
\nonumber
\\
&& \hspace{2cm} \times 
\big[ - 2 ( x+(1-x)r )  \mu (E - (1-x) E_*) + x p^2 \big]^{d/2 -2}.
\label{L1}
\eea
We will need the expansions of these functions to first order in $p^2$:
\begin{subequations}
\bea
L_0(E,p)  &=& r I_1(E) \left[ 1
- \frac{(d-2) p^2}{4 \mu E} \right]+  {\cal O}(p^4),
\label{L0-p}
\\
L_1(E,p)  &=& 
\frac{J_1(E) - I_1(E)}{2 \mu \Delta} 
\left[ 1 + \frac{r \big[4r(1-r) E + (4-4r-d) \Delta \big]  p^2}
           {2 d \mu \Delta^2} \right]
\nonumber \\
&& - \frac{r (d-2) \big[ 4(1-r)E  - d \Delta \big] I_1(E) p^2}
                {8 d \mu^2 E \Delta^2}
+ {\cal O}(p^4),
\label{L1-p}
\eea
\label{L01-p}%
\end{subequations}
where $\Delta=(1-r)E - E_*$.
The expansion for $L_0$ can be obtained most easily by expanding the 
expression in Eq.~\eqref{L0} in powers of $p^2$.
The expansion for $L_1$ can be obtained most easily by 
expanding the integrand in Eq.~\eqref{int-Ln} in powers of $\bm{p}$
and averaging over angles to reduce it to scalar integrals.
The scalar integrals can be expressed in terms of the integrals $J_n$ and $I_n$
given by Eqs.~\eqref{Jn}  and \eqref{In}.
They can all be reduced algebraically to $J_1$ and $I_1$.

\subsection{Two-loop momentum integrals}
\label{sec:2loop}

The two-loop momentum integrals whose integrands have a  $\pi$ propagator
and one or two $D^*$ propagators raised to integer powers are
\bea
K_{lmn}(E) &=& \int_{\bm{p}} \int_{\bm{q}} 
\frac{1}{[ p^2 - 2\mu (E - E_*)]^m [q^2 - 2\mu (E - E_*)]^n} 
\nonumber
\\
&& \hspace{1cm}
\times  \frac{(2\mu)^{-l}}{[ (\bm{p} + \bm{q})^2/(2m) + (p^2 + q^2)/(2M) - E - i \epsilon]^l} .
\label{int-Klmn}
\eea
The momentum integrals with a $\pi$ propagator and a single $D^*$ propagator
can be expressed as integrals over a single Feynman parameter:
\begin{subequations}
\bea
K_{110}(E)  &=& r^{d/2}
\frac{\Lambda^{6-2d} \Gamma(2-d)}{(4 \pi)^d} 
\int_0^1 {\rm d}x\,  (1-x)^{-d/2}
[ 2 \mu (x E_* - E) ]^{d -2},
\label{K10}
\\
K_{120}(E)  &=& r^{d/2}
\frac{\Lambda^{6-2d} \Gamma(3-d)}{(4 \pi)^d} 
\int_0^1 {\rm d}x\,  x (1-x)^{-d/2}
[ 2 \mu (x E_* - E) ]^{d-3},
\label{K20}
\eea
\label{K10,K20}%
\end{subequations}
where $r= \mu_\pi/\mu$.
The momentum integrals with  a $\pi$ propagator and two $D^*$ propagators
can be expressed as an integral over two Feynman parameters:
\bea
K_{111}(E)  &=& r^{d/2}
\frac{\Lambda^{6-2d} \Gamma(3-d)}{(4 \pi)^d} 
\int_0^1 {\rm d}w\,  w [ 2 \mu (w E_* - E) ]^{d-3}
\nonumber
\\
&& \hspace{4cm} 
\times \int_0^1 {\rm d}t\,  
\left[ 1-w + r w^2 t(1-t)\right]^{-d/2}.
\label{K11}
\eea
This function depends on $r$
through the prefactor $r^{d/2}$ and through the integral over $t$.
For a given dimension $d$, $K_{111}(E)$ can be expanded 
in powers of $r$.  The expansion is not straightforward, 
because there are negative powers of $r$
that arise from the $w \to 1$ endpoint region.

\subsection{Poles in $\bm{d-2}$}
\label{sec:poles2}

In a dimensionally regularized loop integral,
poles in $d-2$ are associated with linear ultraviolet (UV) divergences 
in 3 spatial dimensions.  
An UV pole in $d-2$ can arise from a factor of $\Gamma(1-d/2)$ 
or $\Gamma(2-d)$ obtained by integrating over a loop momentum.
In the two-loop momentum  integrals $K_{110}$ and $K_{120}$ defined in 
Eqs.~\eqref{K10,K20}, the integral over the  Feynman parameter
$x$ also gives an UV pole in $d-2$.

The one-loop momentum integrals $J_1$ and $I_1$
given by Eqs.~\eqref{Jn} and \eqref{In} have single poles in $d-2$.
We will need the pole and the constant term:
\begin{subequations}
\bea
J_1(E) &\longrightarrow&
-\frac{\Lambda}{2 \pi}
\left[ \frac{1}{d-2} + \frac12 \log \frac{2 \mu (E_*- E)}{\overline{\Lambda}^2} \right],
\label{J1:d->2}
\\
I_1(E) &\longrightarrow&
-\frac{\Lambda}{2 \pi}
\left[ \frac{1}{d-2} + \frac12  \left( \log \frac{2 \mu E}{ \overline{\Lambda}^2}+ \log r - i \pi \right) \right].
\label{I1:d->2}
\eea
\label{J1I1:d->2}%
\end{subequations}
The momentum scale $\overline{\Lambda}$ 
in the logarithms is 
\beq
\overline{\Lambda} =\sqrt{4 \pi} e^{-\gamma/2} \Lambda,
\label{Lambdabar}
\eeq
where $\gamma$ is Euler's constant.
We will need the constant term in the integral $J_2$
given by Eq.~\eqref{Jn}:
\beq
J_2(E) \longrightarrow
\frac{\Lambda}{8 \pi \mu (E_*- E)}.
\label{J2:d->2}
\eeq
We will also need the pole term in the integral $L_0$
 in Eq.~\eqref{L0}:
\beq
L_0(E,p) \longrightarrow
-\frac{r \Lambda}{2 \pi (d-2)}.
\label{L0:d->2}
\eeq
The two-loop momentum integral $K_{110}$  defined in 
Eq.~\eqref{K10} has a double and single pole in $d-2$,
while $K_{120}$ defined in Eq.~\eqref{K20} has only a single pole:
\begin{subequations}
\bea
K_{110}(E) &\longrightarrow&
\frac{2r \Lambda^2}{(4\pi)^2} 
\left[ \frac{1}{(d-2)^2} 
+ \frac{1}{d-2}  
\left( \log \frac{- 2 \mu (E - E_*)}{\overline{\Lambda}^2} 
+ \frac12 \log r \right) \right],
\label{K10:d->2}
\\
K_{120}(E) &\longrightarrow&
\frac{r \Lambda^2}{(4\pi)^2 (d-2) \mu (E - E_*)} .
\label{K20:d->2}
\eea
\label{K:d->2}%
\end{subequations}

\subsection{Integrals at the $\bm{D^{*0} \bar{D}^0}$ threshold}
\label{sec:threshold}

If the one-loop integral $J_n(E)$ given by Eq.~\eqref{Jn}
is analytically continued to $d=3$ 
and then evaluated at $E = E_*$, it has an infrared divergence for $n \ge 3/2$.
However if it is evaluated at $E = E_*$ and then 
analytically continued to $d=3$, it vanishes
because the integral has no momentum scale:
\beq
J_n(E_*) = 0,
\label{Jn:E*}
\eeq

The two-loop integrals $K_{lmn}(E)$ defined in Eq.~\eqref{int-Klmn} 
can be evaluated analytically at the threshold $E = E_*$.
The integrals $K_{110}(E_*)$, $K_{120}(E_*)$, 
and $K_{111}(E_*)$ each has a single pole in $d-3$.
Their values near $d=3$,
including the pole in $d-3$ and the finite term, are
\begin{subequations}
\bea
K_{110}(E_*) &=& 
 \frac{(-2 )r^{3/2} (2 \mu E_*)}{(4 \pi)^3}
\left( \frac{1}{d-3} - 2 +\frac12 \log r
+\log \frac{2 \mu E_*}{\overline{\Lambda}^2}  -i \pi \right) ,
\label{K10:E*}
\\
K_{120}(E_*) &=& 
\frac{4 r^{3/2} }{(4 \pi)^3}
\left( \frac{1}{d-3} +\frac12 \log r
+\log \frac{2 \mu E_*}{\overline{\Lambda}^2}  -i \pi \right),
\label{K20:E*}
\\
K_{111}(E_*) &=& 
\frac{(-4)r}{(4 \pi)^3}
\bigg[ \frac{\arccos \big(\sqrt{r}\,\big)}{ \sqrt{1-r}} 
\left( \frac{1}{d-3} - 2 + \log r 
+ \log \frac{2 \mu E_*}{\overline{\Lambda}^2}  -i \pi \right) 
\nonumber\\
&& \hspace{2cm} 
+\frac{\rm{d}\ }{{\rm d}d}\,
{}_2F_1(\mbox{$\frac12$}d-1,\mbox{$\frac12$}d-1,\mbox{$\frac12$}d;1-r)
\Big |_{d=3} \bigg] .
\label{K11:E*}
\eea
\end{subequations}
The momentum scale $\overline{\Lambda} $ in the logarithms is 
defined in Eq.~\eqref{Lambdabar}.
The pole in $K_{110}(E_*)$ is the sum of an UV pole
and an infrared pole in $d-3$.

\subsection{Integrals near $\bm{d=3}$}
\label{sec:int3}

The  one-loop integrals $J_1$ and $J_2$ at $d=3$ are
\begin{subequations}
\bea
J_1(E) &=& - \frac{1}{4 \pi} [2 \mu (E_*-E)]^{1/2},
\label{J1:d=3}
\\
J_2(E) &=& - \frac{1}{8 \pi} [2 \mu (E_*-E)]^{-1/2}.
\label{J2:d=3}
\eea
\label{J1J2:d=3}%
\end{subequations}
The  one-loop integrals $I_1$ and $I_2$ at $d=3$ are
\begin{subequations}
\bea
I_1(E) &=& i r^{1/2} \frac{1}{4 \pi} [2 \mu E]^{1/2},
\label{I1:d=3}
\\
I_2(E) &=& - i  r^{-1/2} \frac{1}{8 \pi} [2 \mu E]^{-1/2}.
\label{I2:d=3}
\eea
\label{I1I2:d=3}%
\end{subequations}

The two-loop momentum integrals $K_{lmm}(E) $ near $d=3$
are expressed in terms of
Feynman parameter integrals  in Eqs.~\eqref{K10,K20} and \eqref{K11}.
In order to obtain expressions for these functions
near $d-3$, it is necessary to make subtractions of the integrands 
to make them integrable at $x=1$.  The resulting expression for $K_{110}(E)$,
including the pole in $d-3$ and the finite term, is
\bea
K_{110}(E) &=&
\frac{4r^{3/2}\mu}{(4\pi)^3} 
\Bigg[  \left( \frac{1}{d-3} -2 + \frac12 \log r  
+ \log\frac{2 \mu (E - E_*)}{\overline{\Lambda}^2} - i \pi \right)  (E-2E_*)
\nonumber \\
 &&  \hspace{2cm}
- 2 E_*
- \frac12  \int_0^1 {\rm d}x\, (1-x)^{-3/2}  (E - x E_*)
 \log\frac{E - x E_*}{E - E_*} \Bigg].
\label{K10:d=3}
\eea
The pole with its residue evaluated at $E=E_*$
is also the UV pole in $d-3$ of $K(E_*)$.
The difference between the pole of $K(E_*)$ in Eq.~\eqref{K10:E*} 
and this pole is the infrared pole in $d-3$ of $K(E_*)$.
The UV poles in $d-3$ of $K_{120}(E)$ and $K_{111}(E)$
are the same as the poles in their values at $E_*$ given in 
Eqs.~\eqref{K20:E*} and \eqref{K11:E*}.
The differences between their values at $E$  in Eqs.~\eqref{K20} 
and \eqref{K20} and their values at $E_*$ are finite:
\begin{subequations}
\bea
K_{120}(E) &=& K_{120}(E_*) +
\frac{4 r^{3/2}}{(4\pi)^3} 
\Bigg[ \log\frac{E - E_*}{E_*} 
  - \frac14 \int_0^1 {\rm d}x\, x(1-x)^{-3/2}  \log\frac{E - x E_*}{E - E_*}
 \Bigg],
 \nonumber\\
\label{K20:d=3}
\\
K_{111}(E) &=& K_{111}(E_*)
-\frac{r^{3/2}}{(4\pi)^3}
 \int_0^1 {\rm d}w\, \frac{w}{\sqrt{1-w}(1-w+rw^2/4)}
 \log\frac{E - w E_*}{(1-w)E_*}.
\label{DeltaK11:d=3}
\eea
\end{subequations}

\subsection{Threshold expansions}
\label{sec:threshexp}

Threshold expansions for the two-loop integrals $K_{lmn}(E)$ near $d=3$
cannot be obtained simply by expanding the expressions   
in Eqs.~\eqref{K10:d=3}, \eqref{K20:d=3}, and \eqref{DeltaK11:d=3}
 in powers of $E-E_*$,
because this generates infrared divergences in the Feynman parameter integrals.
The threshold expansion for $K_{111}(E)$ for general $d$
can be obtained in a straightforward way
by expanding the defining integral
in Eq.~\eqref{int-Klmn} in powers of $E-E_*$.
The expansion through second order in $E-E_*$ is
\bea
K_{111}(E) &=& K_{111}(E_*)
+ \left[ K_{211}(E_*) + 2K_{121}(E_*) \right] 2\mu(E - E_*)
\nonumber\\
&&
+ \left[ K_{311}(E_*) + 2 K_{221}(E_*) + K_{122}(E_*)   + 2 K_{131}(E_*)\right]
 [2\mu(E - E_*)]^2
 \nonumber\\
&&
+ {\cal O}\big( (E-E_*)^3\big).
\label{K11:expand}
\eea
The reason the threshold expansion of $K_{111}(E)$
is straightforward is that the only momentum region 
that contributes is where both loop momenta are of order $(\mu |E_*|)^{1/2}$.  

In contrast,  the integrals $K_{110}(E)$ and $K_{120}(E)$ 
have contributions from regions where one loop momentum
is of order $(\mu |E-E_*|)^{1/2}$. 
These contributions have an expansion in non-integer powers of $E-E_*$.
By inserting the alternative expression for the pion propagator
in Eq.~\eqref{piprop} into the defining integral for $K_{lm0}(E)$ 
in Eq.~\eqref{int-Klmn} and then shifting the momentum $\bm{q}$, 
it can be expressed as
\beq
K_{lm0}(E) = r^l
\int_{\bm{p}} \frac{1}{[p^2 - 2 \mu (E-E_*) ]^m} \int_{\bm{q}} 
\frac{1}{ [q^2 - 2 \mu_\pi E_* + r (p^2 -2 \mu (E-E_*)) - i \epsilon]^l}.
\label{Klm0}
\eeq
The expansion of the pion propagator in powers of 
$p^2 -2 \mu (E-E_*)$ produces a sum of
products of one-loop integrals of the form $I_{l+k}(E_*)J_{m-k}(E)$.
With dimensional regularization, the terms with $k \ge m$ 
are zero because the integral over $\bm{p}$ has no scale.
The threshold expansion for $K_{lm0}(E)$ 
is the sum of the terms with the noninteger powers $(E-E_*)^{d/2-m+k}$
and the terms with integer powers
that can be obtained by expanding the 
integrand of Eq.~\eqref{Klm0} in powers of $E-E_*$.
The expansions of $K_{110}(E)$ and $K_{120}(E)$ 
through second order in $E-E_*$ are
\begin{subequations}
\bea
K_{110}(E) &=& K_{110}(E_*) + r I_1(E_*) J_1(E)
+ \left[ K_{210}(E_*) + K_{120}(E_*) \right] 2\mu(E - E_*)
\nonumber\\
&&
+ \left[ K_{310}(E_*) + K_{220}(E_*)  + K_{130}(E_*) \right]
 [2\mu(E - E_*)]^2 + {\cal O}\big( (E-E_*)^3\big) ,
\label{K10:expand}
\\
K_{120}(E) &=& K_{120}(E_*)
 + r I_1(E_*) J_2(E) - r^2 I_2(E_*) J_1(E)
+ \left[ K_{220}(E_*) + 2 K_{130}(E_*) \right] 2\mu(E - E_*)
\nonumber\\
&&
+ \left[ K_{320}(E_*) + 2 K_{230}(E_*)  + 3 K_{140}(E_*) \right]
 [2\mu(E - E_*)]^2 + {\cal O}\big( (E-E_*)^3\big)  .
\label{K20:expand}
\eea
\label{K10,K20:expand}%
\end{subequations}

The integer powers of $E-E_*$ in the threshold expansions in 
Eqs.~\eqref{K11:expand}  and \eqref{K10,K20:expand} 
are determined by the two-loop integrals $K_{lmn}(E_*)$.
By inserting the alternative expression for the pion propagator
in Eq.~\eqref{piprop} into the defining integral for $K_{lmn}(E)$ 
 in Eq.~\eqref{int-Klmn},
its value at the threshold $E=E_*$ reduces to
\beq
K_{lmn}(E_*) = r^l
\int_{\bm{p}} \frac{1}{[p^2]^m} \int_{\bm{q}} 
\frac{1}{[q^2]^n [(\bm{q}+ \sqrt{1-r}\,  \bm{p})^2 + r p^2 -2r \mu E_* - i \epsilon]^l}.
\label{Klmn-E*2}
\eeq
The denominators of the integral over $\bm{q}$
can be combined by introducing a Feynman parameter.  
After evaluating the integral over $\bm{q}$, the integral over $\bm{p}$
can also be evaluated analytically.
The integral over the Feynman parameter gives a hypergeometric function;
\bea
K_{lmn}(E_*) &=& 
\frac{r^{d-m-n}}{(4 \pi)^3}
\left( \frac{1}{4 \pi \Lambda^2} \right)^{d-3}
\frac{\Gamma(l+m+n-d) \Gamma(d/2-m) \Gamma(d/2-n) }
{\Gamma(l) \Gamma^2(d/2)}
\nonumber \\ 
&& \hspace{1cm} \times\,
{}_2F_1(d/2-m,d/2-n,d/2,1-r) \left[ e^{-i \pi}\,  2 \mu E_* \right]^{d-l-m-n}.
\label{KlmnE*2F1}
\eea
If $n=0$, the hypergeometric function reduces to $r^{m-d/2}$.
Inserting the expression for $K_{lmn}(E_*)$ in Eq.\eqref {KlmnE*2F1}
into Eqs.~\eqref{K10:expand}, \eqref{K20:expand}, and \eqref{K11:expand},
the threshold expansions through second order in $E-E_*$ are
\begin{subequations}
\bea
K_{110}(E) &=& K_{110}(E_*)  + r I_1(E_*) J_1(E)
\nonumber\\
&& 
+ \frac{2 r^{3/2} (2\mu E_*)}{(4\pi)^3} 
\left( \frac{1}{d-3}  + 1 + \frac12 \log r 
+ \log\frac{2 \mu E_*}{\overline{\Lambda}^2} - i \pi  \right)  
\frac{E_*-E}{E_*}
\nonumber\\
&& + \frac{r^{3/2} (2 \mu E_*)}{3 (4\pi)^3} 
 \left( \frac{E_*-E}{E_*} \right)^2+ {\cal O}\big( (E-E_*)^3\big),
\label{K10:thresh}
\\
K_{120}(E) &=& K_{120}(E_*)  + r I_1(E_*) J_2(E) - r^2 I_2(E_*) J_1(E)
\nonumber\\
&& 
- \frac{4 r^{3/2}}{3 (4\pi)^3}  \frac{E_*-E}{E_*} 
+ \frac{2 r^{3/2}}{15(4\pi)^3} \left( \frac{E_*-E}{E_*} \right)^2  
+ {\cal O}\big( (E-E_*)^3\big),
\label{K20:thresh}
\\
K_{111}(E) &=& K_{111}(E_*)
+ \frac{4}{(4\pi)^3}  
 \left( (2-r) \frac{\arccos \big(\sqrt{r}\,\big)}{\sqrt{1-r} } + 2 \sqrt{r} \right)
\frac{E_*-E}{E_*} 
\nonumber\\
&& 
+ \frac{2r^{-1}}{(4\pi)^3}  
\bigg( (2-4r+r^2) \frac{\arccos \big(\sqrt{r}\,\big)}{\sqrt{1-r} } +  \frac{2(3-4r)}{3 }\sqrt{r}
\nonumber \\
&& \hspace{1.5cm}
+ 8 \, {}_2F_1\big(\mbox{$-\frac12$},\mbox{$-\frac12$},\mbox{$\frac32$},1-r\big) \bigg)
\left( \frac{E_*-E}{E_*} \right)^2
+ {\cal O}\big( (E-E_*)^3\big).
\label{K11:thresh}
\eea
\label{K:thresh}%
\end{subequations}
The function $\arccos(\sqrt{r}\,)$  in Eq.~\eqref{K11:thresh}
has an expansion in odd powers of $r^{1/2}$:
\beq
\arccos \big(\sqrt{r}\,\big)= 
\frac{\pi}{2} \left( 1 - \frac{2}{\pi} r^{1/2} 
 - \frac{1}{3\pi} r^{3/2}  + \ldots \right) .
\label{arccos-r}
\eeq
The hypergeometric function in Eq.~\eqref{K11:thresh}
has an expansion in powers of $r$:
\beq
{}_2F_1\big(\mbox{$-\frac12$},\mbox{$-\frac12$},\mbox{$\frac32$},1-r\big) = 
\frac{3\pi}{8} \left( 1 - \frac16 r + \frac{1}{24} r^2 + \ldots \right) .
\label{2F1-r}
\eeq

\section{Diagrams for NLO Transition Amplitude}
\label{sec:NLOAmp}

In this Appendix, we give results for the individual diagrams 
that contribute to the transition amplitude $+i{\cal A}^{ij}/2$
for  $D^{*0} \bar D^0 \to D^{*0} \bar D^0$ in the $C=+$ channel,
with the external legs amputated and the initial and final 
$\bar D^0$ on their energy shells.
In the center-of-momentum frame, the amplitude 
${\cal A}^{ij}(E,\bm{p},\bm{p}')$ is a function 
of the total energy $E$ and the relative momenta
$\bm{p}$ and $\bm{p}'$ of the incoming and outgoing charm mesons, respectively.
The LO amplitude is the solution to the Lippmann-Schwinger integral equation
shown in Figure~\ref{fig:LS}. 
The amplitude at NLO can be expressed as
\beq
{\cal A}^{ij}(E,\bm{p},\bm{p}')  = 
(2 \pi/\mu) {\cal A}(E) \delta^{ij} 
+ \sum_n {\cal A}_n^{ij}(E,\bm{p},\bm{p}'),
\label{ALO+NLO}
\eeq
where the sum is over the NLO diagrams 
and the amplitude ${\cal A}(E)$ in the LO term is
\beq
{\cal A}(E)  = 
\frac{\mu/(2 \pi)}{- C_0^{-1} - 2 \mu J_1(E)} .
\label{ALO}
\eeq
The one-loop momentum integral $J_1$ is given in Eq.~\eqref{Jn}.
In the physical dimension $d = 3$, the amplitude ${\cal A}(E)$
reduces to the expression in Eq.~\eqref{A0-Ecm}.
The NLO diagrams for the transition amplitude can be organized 
into five sets of 4 diagrams labeled
$A$, $B$, $C$, and $D$.

\subsection{Pion-exchange diagrams}

\begin{figure}[tb]
\centerline{\includegraphics*[height=9cm,angle=0,clip=true]{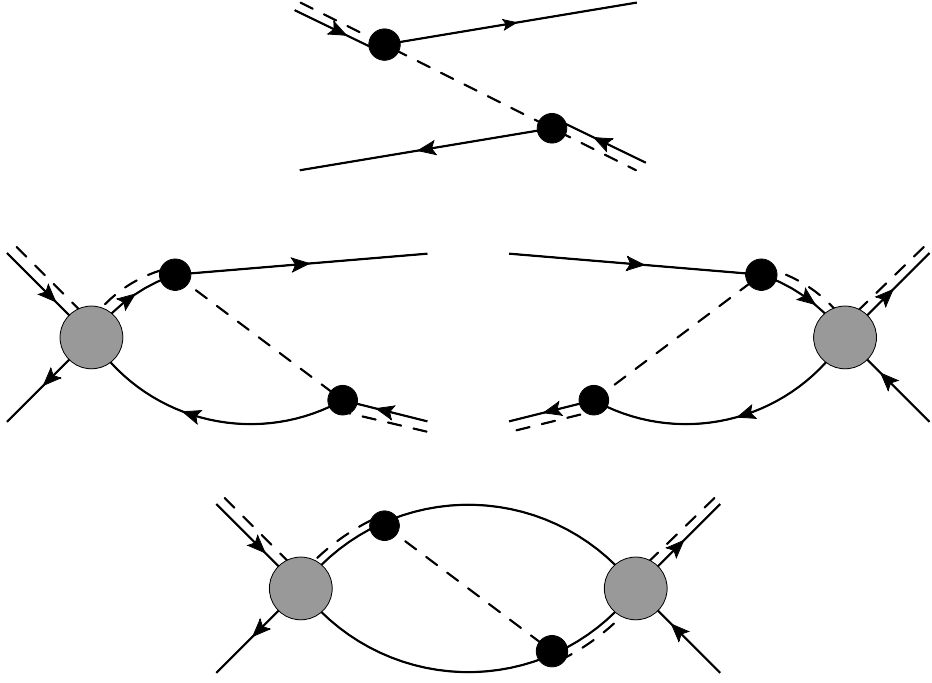}}
\vspace*{0.0cm}
\caption{
Pion-exchange diagrams for $D^{*0} \bar D^0 \to D^{*0} \bar D^0$.
They consist of the tree diagram $A1$,
the one-loop diagrams $A2$ and $A3$ in which either the incoming or the outgoing
charm mesons interact through a LO transition amplitude,
and the two-loop diagram $A4$ in which they both 
interact through LO transition amplitudes.
The absence of arrows on the charm mesons lines in the loop 
for the diagram A4 implies 
a sum over the two possible directions of the arrows.
}
\label{fig:NLOpiex}
\end{figure}

There are four diagrams that involve the emission of a pion 
by $D^{*0}$ and its absorption by $\bar D^0$.
The four pion-exchange diagrams, which are labelled 
$A1$, $A2$, $A3$, and $A4$, are shown in Fig.~\ref{fig:NLOpiex}.
The amplitude for the tree diagram $A1$ is a tensor in the 
vector indices $i$ and $j$ that depends on the relative momenta
$\bm{p}'$ and $\bm{p}$:
\bea
{\cal A}_{A1}^{ij}(E,\bm{p},\bm{p}')  = 
 \left( \frac{g^2}{4m f_\pi^2} \right)
\frac{-2}{E - (p^2 +{p'}^2)/(2M) - (\bm{p} + \bm{p}')^2/(2m) + i \epsilon}
\nonumber
\\
\times \left( \frac{M}{M+m} \bm{p} + \bm{p}' \right)^i
\left( \frac{M}{M+m} \bm{p}' + \bm{p} \right)^j.
\label{A1}
\eea
This can be expressed in a form that depends on the masses
only though reduced masses:
\beq
{\cal A}_{A1}^{ij}(E,\bm{p},\bm{p}')  = 
 \left( \frac{g^2}{4m f_\pi^2} \right)
\frac{(-2)(\sqrt{1-r}\, \bm{p}  + \bm{p}')^i (\sqrt{1-r}\, \bm{p}'  + \bm{p})^j}
       {E - p^2/(2 \mu) - (\bm{p}' + \sqrt{1-r}\, \bm{p})^2/(2\mu_\pi)+ i \epsilon},
\label{A1-mu}
\eeq
where $r = \mu_\pi/\mu$.
Unlike the expression in Eq.~\eqref{A1},
the expression in Eq.~\eqref{A1-mu} is not manifestly invariant
under the simultaneous interchanges 
$\bm{p} \leftrightarrow \bm{p}'$ and $i \leftrightarrow j$.
The amplitude for the diagram $A1$ vanishes in the zero-momentum limit 
$\bm{p},\bm{p}' \to 0$.

The amplitudes for the one-loop diagrams $A2$ and $A3$ in Fig.~\ref{fig:NLOpiex} 
are tensors in the indices $i$ and $j$ that depend on the relative momenta
$\bm{p}'$ and $\bm{p}$, respectively.
They can be decomposed into transverse and longitudinal components.
The Feynman rule for the LO transition amplitude is given in Eq.~\eqref{amprule}.
The amplitudes for the diagrams $A3$ and $A2$ are
\begin{subequations}
\bea
{\cal A}_{A3}^{ij}(E,\bm{p})  &=& 
\left( \frac{g^2}{4m f_\pi^2} \right) \frac{4 \pi \mu^2}{(d-1)\sqrt{1-r}\, p^2}
{\cal A}(E)
\nonumber
\\
&& \times  \Big\{ \big[ - (\Delta(E,p) + r p^2/\mu ) L_0(E,p)
-2 \mu(\Delta(E,p)^2 -2r E_* p^2/\mu ) L_1(E,p)
\nonumber
\\
&& \hspace{4cm}
+ r (\Delta(E,p) + p^2/\mu ) J_1(E) \big]
( \delta^{ij} - p^i p^j/p^2 )
\nonumber
\\
&& \hspace{1cm}
+  (d-1) \big[ \Delta(E,p)  L_0(E,p)
+ 2 \mu \Delta(E,p) (\Delta(E,p) + r p^2/\mu ) L_1(E,p)
\nonumber
\\
&& \hspace{4cm}
- r (\Delta(E,p) - (1-r) p^2/\mu ) J_1(E) \big] p^i p^j/p^2  \Big\} ,
\label{A3}
\\
{\cal A}_{A2}^{ij}(E,\bm{p}')  &=&
{\cal A}_{A3}^{ij}(E,\bm{p} \to \bm{p}')  .
\label{A2}
\eea
\label{A23}%
\end{subequations}
The one-loop integrals $J_1$, $L_0$, and $L_1$ are given by
Eqs.~\eqref{Jn}, \eqref{L0}, and \eqref{L1}, and $\Delta$ 
is a linear function of $E$ and $p^2$:
\beq	
\Delta(E,p) = (1-r)E - E_* - p^2/(2 \mu).
\label{DeltaE}
\eeq
The expression for the amplitude ${\cal A}_{A3}^{ij}(E,\bm{p})$ 
in Eq.~\eqref{A3} has terms proportional to
$\delta^{ij}/p^2$, $p^i p^j/p^4$, and $p^i p^j/p^2$ that are not 
analytic functions of the momentum vector $\bm{p}$
in the neighborhood of $\bm{p}=0$.
The limit $\bm{p} \to 0$ can be obtained by using the  expansions
for $L_0(E,p)$ and $L_1(E,p)$ to first order in $p^2$ given in
Eqs.~\eqref{L01-p}.
The nonanalytic terms cancel in the limit $\bm{p} \to 0$.
The diagram $A2$ has the same zero-momentum limit.
The sum of the amplitudes for the diagrams $A2$ and $A3$
in the zero-momentum limit is
\beq
{\cal A}_{A2+A3}^{ij}(E,0)  = 
\left( \frac{g^2}{4m f_\pi^2} \right) 
\frac{16\pi  r \sqrt{1-r}\,\mu [(E-E_*) J_1(E) -r E I_1(E)]}{d [(1-r)E-E_*]}
{\cal A}(E)\,  \delta^{ij}.
\label{A23:p=0}
\eeq
The one-loop integral $I_1$ is given in Eq.~\eqref{In}.

The two-loop diagram $A4$  in Fig.~\ref{fig:NLOpiex}
is the sum of a diagram in which
a pion is emitted by $D^{*0}$ and absorbed by $\bar D^0$
and a diagram in which
a pion is emitted by $\bar D^{*0}$ and absorbed by $D^0$.
The amplitude for this diagram 
is diagonal in the vector indices $i$ and $j$ and depends only on $E$:
\bea	
{\cal A}^{ij}_{A4}(E)  =
\left( \frac{g^2}{4m f_\pi^2} \right)
\frac{(-16 \pi^2) r \mu}{d \sqrt{1-r}} 
\big[ 2 K_{110}(E) -2 \mu (2 E_* - r E) K_{111}(E)  
\nonumber
\\ 
- (2-r) J_1(E)^2 \big] {\cal A}^2(E) \, \delta^{ij}.
\label{A4}
\eea
The two-loop integrals $K_{110}$ and $K_{111}$ 
are given in Eqs.~\eqref{K10} and \eqref{K11}.

\subsection{$\bm{D^*}$ propagator correction diagrams}

\begin{figure}[tb]
\centerline{\includegraphics*[height=6cm,angle=0,clip=true]{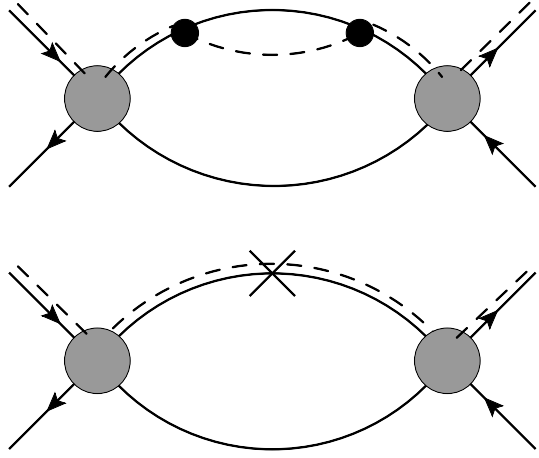}}
\vspace*{0.0cm}
\caption{
$D^*$ propagator correction diagrams for $D^{*0} \bar D^0 \to D^{*0} \bar D^0$.
The two-loop diagram $B1$ has a $D^*$ self-energy subdiagram 
inserted into the $D^*$ propagator. 
The one-loop diagram $B2$ has a $D^*$ self-energy counterterm
inserted into the $D^*$ propagator. 
}
\label{fig:NLOD*se}
\end{figure}

There are two NLO diagrams involving a correction to a $D^*$ propagator.
The two diagrams, which are labelled $B1$ and $B2$, 
are shown in Figure~\ref{fig:NLOD*se}.
The diagram $B1$ has a $D^*$ self-energy subdiagram 
inserted into the $D^*$ propagator.  It can be reduced to
\beq
{\cal A}^{ij}_{B1}(E)  =
\left( \frac{g^2}{4m f_\pi^2} \right)
\frac{(-32 \pi^2) r \mu}{d} \big[ K_{110}(E) - 2 \mu E_* K_{120}(E) \big]
{\cal A}^2(E) \, \delta^{ij}.
\label{B1}
\eeq
The two-loop integrals $K_{110}$ and  $K_{120}$ are given in Eqs.~\eqref{K10,K20}.
The diagram $B2$ has a $D^*$ propagator counterterm 
inserted into the $D^*$ propagator.
The Feynman rule for the self-energy counterterm 
is given in Eq.~\eqref{D*counterterm}.
With the complex on-shell renormalization scheme for the $D^{*0}$ propagator,
the diagram $B2$ can be reduced to
\beq
{\cal A}^{ij}_{B2}(E)  =
\left( \frac{g^2}{4m f_\pi^2} \right)
\frac{(-16 \pi^2) r^2 \mu}{d} I_1(E_*) \big[ 4 \mu E_* J_2(E) - d J_1(E)  \big]
{\cal A}^2(E) \, \delta^{ij}.
\label{B2}
\eeq
The one-loop integrals $J_n$ are given in Eq.~\eqref{Jn}.

\subsection{$\bm{\nabla^2}$ vertex diagrams}

\begin{figure}[tb]
\centerline{\includegraphics*[height=9cm,angle=0,clip=true]{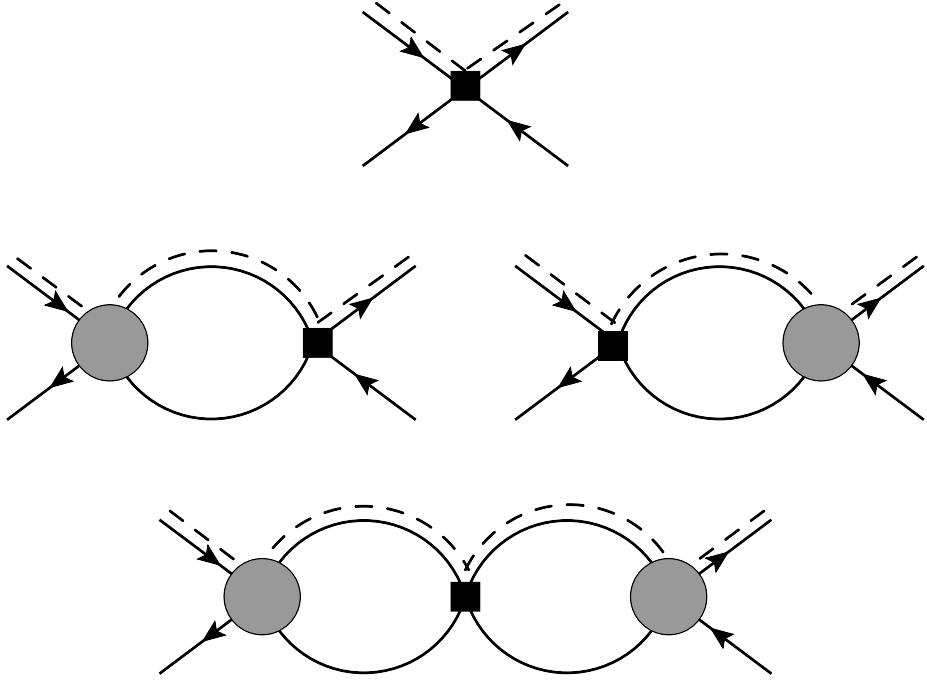}}
\vspace*{0.0cm}
\caption{
The $\nabla^2$ vertex diagrams for $D^{*0} \bar D^0 \to D^{*0} \bar D^0$.
They consist of the $\nabla^2$ vertex $C1$,
the one-loop diagrams $C2$ and $C3$ in which either the incoming or the outgoing
charm mesons interact through a LO transition amplitude,
and the two-loop diagram $C4$ in which they both
interact through LO transition amplitudes.
}
\label{fig:NLOnabla2}
\end{figure}

There are four NLO diagrams with a $\nabla^2$ vertex.
The four diagrams, which are labelled $C1$, $C2$, $C3$, and $C4$,
are shown in Fig.~\ref{fig:NLOnabla2}.
The amplitude for the first diagram $C1$ is just the $\nabla^2$ vertex
in Eq.~\eqref{vertex:grad^2}:
\beq	
{\cal A}_{C1}^{ij}(E,p,p')  = 
(-C_2/2) \left( p^2 + {p'}^2 \right) \delta^{ij} .
\label{C1}
\eeq
The factor of $p^2 + {p'}^2$ depends on the relative momenta
of the incoming and outgoing charm mesons.
If there is a contact interaction between the incoming charm mesons, 
$p$ is replaced by a loop momentum $k$.
The amplitude for the diagram $C2$ in which the incoming 
charm mesons interact through a LO transition amplitude
can be obtained from Eq.~\eqref{C1}
by multiplying it by $(4\pi/\mu) J_1(E) {\cal A}(E)$
and replacing $p^2$ by the loop-integral-weighted average of $k^2$:
\beq	
{\cal A}_{C2}^{ij}(E,p')  = 
- 2 \pi C_2 J_1(E) {\cal A}(E)
\left( \langle k^2 \rangle + {p'}^2 \right) \delta^{ij} .
\label{C2}
\eeq
The loop-integral-weighted average of $k^2$ is defined by
\beq	
\langle k^2 \rangle = 
\frac{\int_{\bm{k}} k^2/[E - E_* - k^2/(2\mu)]}
       {\int_{\bm{k}} 1/[E - E_* - k^2/(2\mu)]}.
\label{<k^2>-int}
\eeq
With dimensional regularization, 
the loop-integral-weighted average is very simple:
\beq	
\langle k^2 \rangle = 2\mu (E - E_*).
\label{<k^2>}
\eeq
The amplitude for the diagram $C3$ is obtained from 
Eq.~\eqref{C2} by replacing $p'$ by $p$.
The amplitude for the diagram $C4$ is
\beq	
{\cal A}_{C4}^{ij}(E)  = 
- 16 \pi^2 C_2 J_1(E)^2 {\cal A}^2(E) \langle k^2 \rangle  \delta^{ij} .
\label{C4}
\eeq
The sum of the four diagrams $C1$, $C2$, $C3$, and $C4$ 
has a multiplicative factor $1 + 4 \pi J_1 {\cal A}$ that can be simplified by using 
the expression for $A_0(E)$ in Eq.~\eqref{ALO}:
\beq
{\cal A}_{C}^{ij}(E,p,p')  = 
\frac{\pi C_2}{\mu C_0} \Big[ (p^2 + {p'}^2) {\cal A}(E) 
+ 16 \pi \mu (E - E_*) J_1(E) {\cal A}^2(E)  \Big] \delta^{ij}.
\label{C1234}
\eeq

\subsection{$\bm{D^* \bar{D}}$ counterterm diagrams}

There are four NLO diagrams that can be obtained
from the diagrams in Figure~\ref{fig:NLOnabla2}
by replacing the $\nabla^2$ vertex by a $D^* \bar D$ counterterm.
The four diagrams are labelled $D1$, $D2$, $D3$, and $D4$.  
The amplitude for  the diagram $D1$ is just the counterterm vertex
in Eq.~\eqref{vertex:counterterm}:
\beq	
{\cal A}_{D1}^{ij}  = -[ \delta C_0 + \delta D_0 E]  \, \delta^{ij} .
\label{D1}
\eeq
The amplitudes for the diagrams $D2$ and $D3$ each
differs from this by a multiplicative factor 
$4 \pi J_1 {\cal A}$. The amplitude for the diagram $D4$
differs by two such factors. 
The sum of the four diagrams $D1$, $D2$, $D3$, and $D4$
therefore has a factor $[1 + 4 \pi J_1 {\cal A}]^2$ 
that can be simplified by using 
the expression for $A_0(E)$ in Eq.~\eqref{ALO}:
\beq	
{\cal A}^{ij}_{D}(E)  =
-\frac{4 \pi^2 [\delta C_0 +\delta D_0 E]}{\mu^2 C_0^2}  \, 
{\cal A}^2(E) \delta^{ij}.
\label{D1234}
\eeq
The $\delta C_0$ term in this expression can also be obtained 
from the LO transition amplitude 
in Eq.~\eqref{ALO} by replacing $C_0^{-1}$ by $C_0^{-1} - \delta C_0/C_0^2$
and expanding the amplitude to first order in $\delta C_0$.

\subsection{Complete NLO amplitude}

The complete NLO term in the transition amplitude 
${\cal A}^{ij}(E,\bm{p},\bm{p}')$ is the sum of
(A) the pion-exchange diagrams in 
Eqs.~\eqref{A1}, \eqref{A23}, and \eqref{A4},
(B) the $D^*$ propagator insertion diagrams in Eqs.~\eqref{B1} and \eqref{B2},
(C) the $\nabla^2$ vertex diagrams in Eqs.~\eqref{C1234},
and (D) the $D^* \bar D$ counterterm diagrams in Eq.~\eqref{D1234}.
In the zero-momentum limit $\bm{p},\bm{p}' \to 0$,
the pion-exchange diagrams reduce to the sum of  
Eqs.~\eqref{A23:p=0} and \eqref{A4}
and the $\nabla^2$ vertex diagrams in Eqs.~\eqref{C1234}
reduce to the single term with the factor of ${\cal A}^2(E)$.

\subsection{Poles in $\bm{d-2}$}

The momentum integrals with poles in $d-2$ are given in 
Section~\ref{sec:poles2}.
In the NLO term in the transition amplitude 
${\cal A}^{ij}(E,\bm{p},\bm{p}')$, all the terms with a double pole in $d-2$
have the tensor structure $\delta^{ij}$.  All the terms 
with a single pole in $d-2$ also have the tensor structure $\delta^{ij}$,
with the exception of terms from the diagram $A3$
with the tensor structure  $p^ip^j/p^2$
and terms from the diagram $A2$
with the tensor structure  ${p'}^i {p'}^j/{p'}^2$.
Upon using the identities $L_0 = r J_1$ and $L_1=0$ for the pole terms, 
the diagram $A3$  in Eq.~\eqref{A3} reduces to
\beq
{\cal A}^{ij}_{A3}(E,\bm{p})  \longrightarrow
\left( \frac{g^2}{4m f_\pi^2} \right)
\frac{4 \pi r \sqrt{1-r} \, \mu}{d-1} J_1
{\cal A}(E)\,  \big[ \delta^{ij} + (d-2) p^i p^j /p^2 \big].
\eeq
Thus the single pole in $d-2$ has the tensor structure  $\delta^{ij}$
even at nonzero momentum.

We now consider the terms with the tensor structure $\delta^{ij}$
in the zero-momentum limit.
The pion-exchange diagrams $A2$ and $A3$  in Eq.~\eqref{A23:p=0}
and the $\nabla^2$ vertex diagrams $C$ in Eq.~\eqref{C1234}
have single poles in $d-2$:
\begin{subequations}
\bea
{\cal A}_{A2+A3}^{ij}(E,0)  &\longrightarrow&
\left( \frac{g^2}{4m f_\pi^2} \right) 
\frac{(-4) r \sqrt{1-r}\,  \mu \Lambda}{d-2}
{\cal A}(E)\,  \delta^{ij},
\label{A23:d->2}
\\
{\cal A}_{C}^{ij}(E,0,0)  &\longrightarrow&
\frac{8 \pi C_2 \Lambda}{(d-2)C_0} (E_* - E )  {\cal A}^2(E) \delta^{ij}.
\label{C:d->2}
\eea
\label{A23C:d->2}%
\end{subequations}
The two-loop pion-exchange diagram $A4$  in Eq.~\eqref{A4}
and the sum of the $D^*$ propagator correction diagrams $B1$ and $B2$
in Eqs.~\eqref{B1} and \eqref{B2}
have double and single poles in $d-2$:
\begin{subequations}
\bea	
{\cal A}^{ij}_{A4}(E) & \longrightarrow&
\left( \frac{g^2}{4m f_\pi^2} \right)
4 r  \sqrt{1-r}\, \mu \Lambda^2
\left[  \frac{1}{(d-2)^2} \right.
\nonumber \\
&& \hspace{1cm}
\left. + \frac{1}{d-2}
\left( \log \frac{2 \mu (E_* - E)}{\overline{\Lambda}^2} 
- \frac{r \log r}{4 (1-r)} - \frac12 \right) \right]
 {\cal A}^2(E) \, \delta^{ij},
\label{A4:d->2}
\\
{\cal A}^{ij}_{B}(E) & \longrightarrow&
\left( \frac{g^2}{4m f_\pi^2} \right)
2 r ^2 \mu \Lambda^2
\left[ \frac{1}{(d-2)^2} \right.
\nonumber \\
&&  \hspace{1cm}
\left. + \frac{1}{d-2}
\left( \log \frac{2 \mu E_*}{\overline{\Lambda}^2} 
+\frac12  \log r + \frac12 -i \pi \right) \right]
 {\cal A}^2(E) \, \delta^{ij}.
\label{B:d->2}
\eea
\label{A4B:d->2}
\end{subequations}

\subsection{Poles in $\bm{d-3}$}

The only NLO diagrams for the transition amplitude that have 
poles in $d-3$ are the two-loop pion-exchange diagram $A4$
in Eq.~\eqref{A4} and the $D^*$ self-energy insertion diagram $B1$
in Eq.~\eqref{B1}.
The momentum integrals with poles in $d-3$ are 
the two-loop integrals $K_{110}$, $K_{120}$, and $K_{111}$.
The poles are given in Eqs.~\eqref{K10:d=3}, \eqref{K20:E*}, 
and  \eqref{K11:E*}, respectively.
The pole terms are
\begin{subequations}
\bea
{\cal A}^{ij}_{A4}(E)  &\longrightarrow&
\left( \frac{g^2}{4m f_\pi^2} \right)
\frac{(-2) r^2 \mu^3}{3 \pi (d-3)} 
\left[ \frac{r^{1/2}}{ \sqrt{1-r}} (E-2E_*)  \right.
\nonumber \\
&& \hspace{4cm} \left.
+ \frac{\arccos(\sqrt{r}\,)}{1-r} (2 E_* - r E) \right] 
{\cal A}^2(E) \, \delta^{ij},
\label{A4:d->3}
\\
{\cal A}^{ij}_B(E)  &\longrightarrow&
\left( \frac{g^2}{4m f_\pi^2} \right)
\frac{(-2)  r^{5/2} \mu^3}{3 \pi (d-3)} (E + E_*)  
{\cal A}^2(E) \, \delta^{ij}.
\label{B1:d->3}
\eea
\label{A4B1:d->3}%
\end{subequations}

\end{appendix}

\end{document}